\documentclass[preprint2]{aastex}

\shorttitle{Magnetically Supported Accretion Disks}
\shortauthors{Oda et al.}

\begin{document}


\title{Thermal Equilibria of Magnetically Supported, Black Hole 
Accretion Disks}


\author{H. Oda\altaffilmark{1}, M. Machida\altaffilmark{2},
  K.E. Nakamura\altaffilmark{3} and R. Matsumoto\altaffilmark{4}}


\altaffiltext{1}{Graduate School of Science and Technology, Chiba University,
  1-33 Yayoi-cho, Inage-ku, Chiba 263-8522, Japan; oda@astro.s.chiba-u.ac.jp}
\altaffiltext{2}{Division of Theoretical Astronomy, National Astronomical
  Observatory of Japan, 2-21-1 Osawa, Mitaka, Tokyo 181-8588, Japan}
\altaffiltext{3}{Department of Sciences, Matsue National College of
  Technology, 14-4 Nishiikuma-cho, Matsue, Shimane 690-8515, Japan}
\altaffiltext{4}{Department of Physics, Graduate School of Science,
  Chiba University, 1-33 Yayoi-cho, Inage-ku, Chiba 263-8522, Japan}


\begin{abstract}
We present new thermal equilibrium solutions for optically thin 
and optically thick disks incorporating magnetic fields. The purpose of 
this paper is to explain the bright hard state and the bright/slow 
transition observed in the rising 
phases of outbursts in black hole candidates. On the
basis of the results of three-dimensional MHD simulations, we assume
that magnetic
fields inside the disk are turbulent and dominated by the azimuthal
component and that the azimuthally averaged Maxwell
stress is proportional to the total (gas, radiation, and magnetic)
pressure. We
prescribe the magnetic flux advection rate to
determine the azimuthal magnetic flux at a given radius. Local thermal
equilibrium solutions are obtained by equating the heating,
radiative cooling, and heat advection terms.

We find magnetically supported ($\beta = \left( p_{\rm gas} +
p_{\rm rad}\right) / p_{\rm 
mag} < 1$), thermally stable solutions for
both optically thin disks and optically thick disks, in which the
heating enhanced by the strong magnetic field balances the
radiative cooling. The temperature in a low-$\beta$ disk ($T \sim
10^{7-11} {\rm K}$) is
lower than that in an advection dominated accretion flow (or radiatively
inefficient accretion flow) but higher than that in a 
standard disk. We also study the radial dependence of the thermal
equilibrium solutions.

The optically thin, low-$\beta$ branch extends to $ \dot
M \gtrsim 0.1 {\dot M}_{\rm Edd}$ where $\dot M$ is the mass accretion
rate and ${\dot M}_{\rm Edd}$ is the Eddington mass accretion rate, in
which the temperature anti-correlates with the mass accretion rate. Thus
optically thin low-$\beta$ disks can explain the bright hard
state. Optically thick,
low-$\beta$ disks have the radial dependence of the effective
temperature $T_{\rm eff} \propto
\varpi^{-3/4}$. Such disks will be observed as staying in a high/soft 
state.
Furthermore, limit cycle oscillations between an 
optically thick low-$\beta$ disk
and a slim disk will occur because the optically thick
low-$\beta$ branch intersects with the radiation pressure dominated
standard disk branch. These limit cycle oscillations will show a smaller
luminosity variation than that between a standard disk and
a slim disk.
\end{abstract}





\section{Introduction}
X-ray spectral states of galactic black hole
candidates (BHCs) have been classified on the basis of their spectral shape and
X-ray luminosity. In the high/soft state, the X-ray spectrum is dominated
by a modified
blackbody (thermal) component with characteristic temperature $\sim
1~ {\rm keV}$.
The low/hard state is observed at low luminosity ($L < 0.1
L_{\rm Edd}$), where
$L_{\rm Edd} = 4 \pi c G M / \kappa_{\rm es} \sim 1.47 \times 10^{39}
(M/10M_\sun)(\kappa_{\rm es}/0.34 ~ {\rm cm}^{2} {\rm g}^{-1})^{-1} ~ {\rm erg} ~ {\rm s}^{-1} $ is
the Eddington luminosity. Here 
$M$ is the black hole mass and $\kappa_{\rm es}$ is the electron
scattering opacity. The X-ray spectrum in the low/hard
state is dominated by a power-law component (photon index $\Gamma \sim 1.7$)
with an exponential cutoff at $\sim 200 ~ {\rm keV}$.
The very high/steep power law state is observed at high luminosity
($\sim L_{\rm Edd}$), in
which the blackbody component and the power-law component become
comparable. 
Super critical accretion flows have been identified by some authors in
narrow-line Seyfert 1 galaxies \citep[NLS1s; e.g.,][]{kawa03} and in
ultraluminous X-ray sources \citep[ULXs; e.g.,][\citeyear{vier08}]{vier06}.
The spectral fitting models used in their paper are based on the
slim disk model \citep[e.g.,][]{abra88}. These objects are referred to
as staying in the slim disk state.

Recently, \cite{miya08} analyzed the results of RXTE observations of the
black hole candidate GX 339-4
in the rising phases of transient outbursts. When
the luminosity is lower than $\sim 0.07 L_{\rm Edd}$, the X-ray
spectrum is dominated by a power-law component with an exponential
cutoff, and the cutoff
energy is roughly constant $\sim 200 {\rm keV}$. This is a
characteristic feature of
the low/hard state. By contrast, when the luminosity becomes higher than
$\sim 0.07 L_{\rm Edd}$, the
cutoff energy anti-correlates with the luminosity and decreases to $\sim 50
{\rm keV}$.
This suggests that
the electron temperature of a disk decreases as the luminosity
increases. Furthermore, this state has been observed up to $\sim 0.3
L_{\rm Edd}$. \cite{miya08} have named it the ``bright hard state''.

\cite{gier06} have reported
two types of hard-to-soft transitions. One is the bright/slow
transition that occurs at
$\sim 0.3 L_{\rm Edd}$ and takes more than 30 days. The other is
the dark/fast transition which occurs at less than $0.1 L_{\rm Edd}$ and
takes less than 15 days \citep[see also][]{done03}.

In the conventional theory of accretion disks, the concept of phenomenological
$\alpha$-viscosity is introduced. In this framework, the $\varpi
\varphi$-component of the stress tensor is assumed to be proportional to
the sum of the gas pressure $p_{\rm gas}$ and
radiation pressure $p_{\rm rad}$, $t_{\varpi \varphi} = \alpha_{\rm
SS} (p_{\rm gas}+p_{\rm rad})$. Here $\alpha_{\rm SS}$ is the viscosity
parameter
introduced by \cite{shak73}.
The standard disk model was developed by
\cite{shak73}, under the assumption that a substantial fraction of the viscously
dissipated energy in an optically thick disk is radiated locally. This
standard disk model reproduced the thermal component
in the high/soft state but cannot explain the hard X-ray spectrum observed
in the low/hard state.

\cite{thor75} proposed that hard X-rays from Cyg X-1 are produced in an inner
optically thin hot disk. \cite{shib75} studied the structure and
stability of optically thin hot accretion disks. \cite{eard75} and
\citet[][hereafter SLE]{shap76} constructed a model for optically thin
two-temperature
accretion disks in which the ion temperature is higher than the
electron temperature. However, these disks are thermally unstable.

\cite{ichi77} pointed out the importance of energy advection in hot,
magnetized accretion flows, and obtained steady solutions
of optically thin disks. In this model, the viscously dissipated energy is
stored in the gas as entropy and advected into the central object
rather than being radiated. Such geometrically thick, optically thin,
advection-dominated accretion flows
(ADAFs) or radiatively inefficient accretion flows
(RIAFs) have been studied
extensively by \citet[][\citeyear{nara95}]{nara94}, and
\cite{abra95}. 
\cite{esin97} found that the maximum luminosity of the ADAF/RIAF
solutions is $\sim 0.4 \alpha^2 L_{\rm Edd}$. These solutions explain
the low/hard state.

However, the ADAF/RIAF solutions cannot explain the high luminosity
observed in the bright hard state and during the bright/slow transition
unless $\alpha \sim 1$. Furthermore, the ADAF/RIAF solutions do not show
an anti-correlation between the luminosity and the electron
temperature (therefore, the cutoff energy) observed in the bright hard state.

In the light of this discussion, our main aim is to propose steady
models of magnetically supported black hole
accretion disks that can explain the bright hard state and the
bright/slow transition.

\cite{okad89} constructed a model of non-relativistic equilibrium tori
with purely azimuthal magnetic fields in the pseudo-Newtonian potential
\citep{pacz80}, which has been used as initial conditions of numerical
simulations \citep[e.g.,][]{mach00,mach03,mach06}. \cite{komi06}
developed a model of relativistic equilibrium tori with strong toroidal
magnetic fields around a rotating black hole. However, the angular
momentum transport was not taken into account in their models.

The origin of the efficient angular momentum transport (that is,
$\alpha$-viscosity) in accretion
disks had been a puzzle until \cite{balb91} pointed out the importance
of the magneto-rotational instability (MRI). The MRI can excite and
maintain magnetic turbulence. The Maxwell stress generated by 
turbulent magnetic
fields dominates the total stress and efficiently transports the angular
momentum of the disk material. Many researchers have studied the
growth and saturation of the MRI by conducting local
and global magnetohydrodynamic (MHD) simulations. 

Global three-dimensional MHD simulations of
optically thin, radiatively inefficient accretion disks indicate that the
amplification of magnetic fields becomes saturated when $\beta = p_{\rm gas
}/ p_{\rm mag} \sim 10$,
and that then the disk approaches a quasi-steady
state. In this quasi-steady state, the ratio of the azimuthally averaged
Maxwell stress to the azimuthally averaged gas pressure, 
 which corresponds to
the $\alpha$-parameter in the conventional theory of accretion disks,
is $\sim 0.01-0.1$
\citep[e.g.,][]{hawl00,mach00,hawl01,mach03}. 

Three-dimensional MHD simulations have
revealed that magnetic fields contribute to not only the angular
momentum transport but also to disk heating
 via the dissipation of magnetic energy
inside the disk \citep[e.g.,][]{mach06,hiro06,krol07}.

\cite{mach06}
found that when the mass
accretion rate exceeds the threshold for the onset of a cooling
instability, an optically thin, radiatively inefficient,
hot, gas pressure dominated accretion disk (ADAF/RIAF-like disk)
evolves towards an optically thin, radiatively efficient, cool, magnetic pressure
dominated accretion disk (low-$\beta$ disk). 
In their simulation, when the density of the accretion disk exceeds a
critical density, the cooling instability taking place in the
ADAF/RIAF-like disk results in a decrease in the gas pressure. Thus
the disk shrinks in the vertical direction. During that time, the
azimuthal magnetic flux $\langle B_{\varphi} \rangle H$ is almost
conserved
because the cooling
time scale is faster than that of the buoyant escape of the magnetic
flux, where $\langle B_{\varphi} \rangle$ is 
the mean azimuthal magnetic field, and $H$ is the half thickness of
the disk. As the disk shrinks, the magnetic
pressure exceeds the gas pressure because the magnetic pressure
increases due to flux conservation.
Then the disk is supported by magnetic pressure and stops shrinking. In such
low-$\beta$ disks, the magnetic flux
cannot escape through the Parker instability \citep{park66} because the
strong magnetic tension suppresses the
growth of the Parker instability \citep[see][]{shib90}. Therefore the
disk stays in a quasi-equilibrium cool state.

In this state, the dissipated magnetic
energy is mainly radiated (not advected), and the Maxwell stress is
proportional to the sum of the gas pressure 
and the magnetic pressure. The luminosity of the
disk becomes higher than that of the initial ADAF/RIAF state. These
numerical 
results indicate that a luminous hard state exists in BHCs.

\cite{oda07} constructed a steady model of optically thin accretion
disks on the basis of these results from three-dimensional MHD simulations
of accretion disks. They found
that when the mass accretion rate exceeds the threshold for the onset of
thermal instability, the thermally and viscously stable low-$\beta$ branch
 appears. These solutions confirmed the results by \cite{mach06}. 
On this branch, the luminosity emitted by an accretion disk can exceed $0.2
 L_{\rm Edd}$. Furthermore, the mean temperature is lower than that of the
ADAF/RIAF (higher than that of the standard disk) and anti-correlates
with the mass accretion rate. 
Therefore, they concluded that optically thin low-$\beta$ disks can
explain the bright/hard state.

In this paper, we extend the model of optically thin low-$\beta$ disks
to the optically thick regime. We obtain local thermal equilibrium curves
of accretion disks incorporating the magnetic fields.
The total pressure is the sum of the gas
pressure, the radiation pressure, and the magnetic pressure. We consider
the thermal bremsstrahlung radiation in the optically thin limit and
blackbody radiation in the optically thick limit as radiative
cooling. To complete the set of basic
equations, we specify the advection rate of the azimuthal magnetic
flux. The basic equations are presented in $\S$
\ref{model}. In $\S$ \ref{result} we obtain the thermal equilibrium
curves and study their radial dependence. $\S$ \ref{discussion} is
devoted to a discussion. We summarize the paper in $\S$ \ref{summary}

 \section{Models and Assumptions} \label{model}

 \subsection{Basic Equations}
We extended the basic equations for one-dimensional steady black hole
accretion flows \cite[e.g.,][]{kato08}, incorporating magnetic
fields. We adopt cylindrical coordinates
$(\varpi,\varphi,z)$. General relativistic effects are
simulated using the pseudo-Newtonian potential $\psi =
-GM/(r-r_{\rm s})$ \citep{pacz80}, where $G$ is the
gravitational constant, $M$ is the black hole mass (we assume $M
= 10 M_\sun$ in this paper), $r = (\varpi^2 + z^2)^{1/2}$, and $r_{\rm s} =
2GM/c^2 $ is the Schwarzschild radius.

We start with the resistive MHD equations
\begin{eqnarray}
 \frac{\partial \rho}{\partial t} + \nabla \cdot \left( \rho
 \mbox{\boldmath{$v$}} \right) = 0 ~,
 \label{eq:vec_con}
\end{eqnarray}
\begin{eqnarray}
 \rho \left[ \frac{\partial \mbox{\boldmath{$v$}}}{\partial t} + \left(
 \mbox{\boldmath{$v$}} 
 \cdot \nabla \right) \mbox{\boldmath{$v$}}
   \right] &=& \nonumber \\
 - \rho \nabla \psi - \nabla \left( p_{\rm gas} + p_{\rm rad}
   \right) &+& \frac{\mbox{\boldmath{$j$}} \times
\mbox{\boldmath{$B$}}}{c} ~,
 \label{eq:vec_mom}
\end{eqnarray}
\begin{eqnarray}
 \frac{\partial \left( \rho \epsilon \right)}{\partial t} &+& \nabla \cdot
 \left[ \left(
 \rho \epsilon + p_{\rm gas} + p_{\rm rad}\right) \mbox{\boldmath{$v$}}
 \right] \nonumber \\
&-& \left(
 \mbox{\boldmath{$v$}} \cdot \nabla \right) \left( p_{\rm gas}+p_{\rm rad}
 \right) = q^{+} - q^{-} ~,
 \label{eq:vec_ene}
\end{eqnarray}
\begin{eqnarray}
 \frac{\partial \mbox{\boldmath{$B$}}}{\partial t} = \nabla \times
  \left( \mbox{\boldmath{$v$}}
 \times \mbox{\boldmath{$B$}} - \frac{4 \pi}{c}\eta
 \mbox{\boldmath{$j$}} \right) ~,
 \label{eq:vec_ind}
\end{eqnarray}
where $\rho$ is the density, $\mbox{\boldmath{$v$}}$ is the velocity,
$\mbox{\boldmath{$B$}}$
is the magnetic field, $\mbox{\boldmath{$j$}} = c \nabla \times
\mbox{\boldmath{$B$}} / 4 \pi$
is the current density, and $\rho \epsilon = p_{\rm gas} / \left(\gamma
-1\right) + 3 p_{\rm rad}$ is the internal energy. Here, $\gamma = 5/3$
is the specific heat ratio and $p_{\rm gas} =
\mathcal{R} \rho T / \mu$ is the gas pressure where $\mathcal{R}$ is
the gas constant, $\mu$ is the mean molecular weight (we assumed to be
$0.617$), and $T$ is the temperature. In the energy equation
(\ref{eq:vec_ene}), $q^{+}$ is the heating rate, and $q^{-}$ is the
radiative cooling rate. In the induction equation (\ref{eq:vec_ind}),
$\eta \equiv c^2/4 \pi \sigma_{\rm c}$ is the magnetic diffusivity, where
$\sigma_{\rm c}$ is the electric conductivity.

 \subsubsection{Azimuthally Averaged Equations}
\begin{figure}
 \epsscale{1.0}
 \plotone{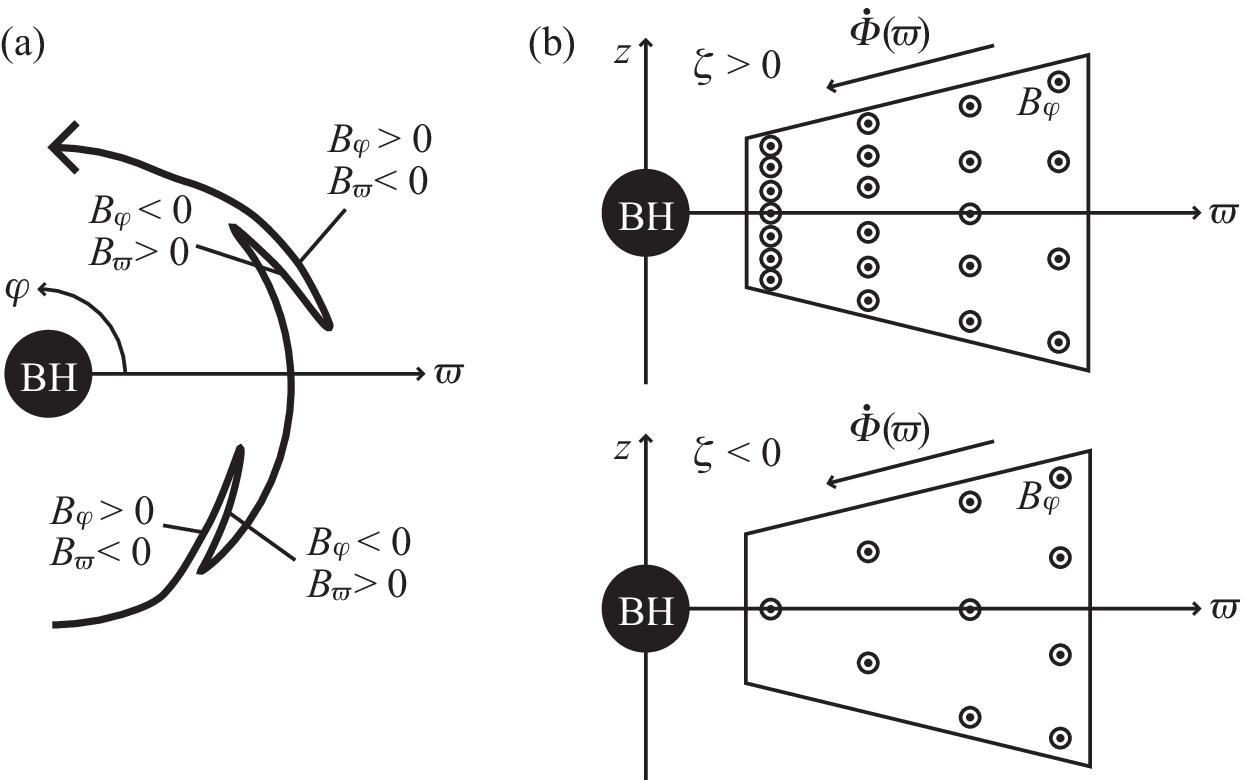}
 \caption{ Schematic picture of magnetic field lines inside the
 accretion disk. (a) The magnetic fields are decomposed into
 the mean azimuthal magnetic fields and fluctuating fields. The
 azimuthal average of the radial component of the magnetic fields,
 $\langle B_{\varpi} \rangle$, is small because positive and negative
 $B_{\varpi}$ cancel out. However, the azimuthal average of
 the product of the radial and azimuthal component $\langle B_{\varpi}
 B_{\varphi} \rangle $ has a large negative value because $B_{\varpi}
 B_{\varphi}$ does not change sign when magnetic fields are deformed by
 nonlinear growth of the MRI. (b) The advection of the azimuthal magnetic
 flux. The radial dependence of the magnetic flux
 advection rate $\dot \Phi$ is parameterized by $\zeta$. When $\zeta >
 0$, $\dot \Phi$ increases with a decreasing radius. However, $\dot
 \Phi$ decreases when $\zeta < 0$. The mean azimuthal
 magnetic flux is conserved in the radial direction when $\zeta =
 0$. \label{line}} 
\end{figure}

Three-dimensional global MHD and local Radiation-MHD simulations of
black hole accretion
disks showed that magnetic fields inside the disk are turbulent and
dominated by the azimuthal component in a quasi-steady state
\cite[e.g.,][]{mach06, hiro06}. On the basis of results of these
simulations, we decomposed the magnetic fields into the
mean fields $\mbox{\boldmath{$\bar{B}$}} =
\left( 0, \langle B_{\varphi} \rangle, 0 \right)$ and 
fluctuating fields $\delta \mbox{\boldmath{$B$}} = \left( \delta B_{\varpi},
 \delta B_{\varphi}, \delta B_{z} \right)$ and also decomposed the
velocity into the mean velocity $\mbox{\boldmath{$\bar{v}$}} =
(v_{\varpi}, v_{\varphi}, v_{z})$ and the
fluctuating velocity $\delta \mbox{\boldmath{$v$}} = \left(\delta v_{\varpi},
 \delta v_{\varphi}, \delta v_{z} \right)$. Here $\langle ~~ \rangle$
denotes the azimuthal average. We assume that the
fluctuating components vanish when azimuthally averaged, $\langle \delta
\mbox{\boldmath{$v$}} \rangle = \langle \delta \mbox{\boldmath{$B$}}
\rangle = 0$, and that the 
radial and vertical components of the magnetic fields are negligible
compared with that of the azimuthal component, $|\langle B_{\varphi}
\rangle + \delta B_{\varphi}| \gg |\delta B_{\varpi}|$, $|\delta
B_{z}|$ (see panel (a) in figure \ref{line}).

Let us derive the azimuthally averaged equations. We assume that the disk is
 in a steady state and in hydrostatic balance in the vertical
direction. By azimuthally averaging equations (\ref{eq:vec_con}) -
(\ref{eq:vec_ind}) and ignoring the second order terms of $\delta
 \mbox{\boldmath{$v$}}$, $\delta B_{\varpi}$,
 and $\delta B_{z}$, we obtain
\begin{eqnarray}
 \label{eq:con}
 \frac{1}{\varpi} \frac{\partial}{\partial \varpi} \left( \varpi \rho
 v_{\varpi} \right) +
 \frac{\partial}{\partial z} \left( \rho v_{z} \right) = 0 ~,
\end{eqnarray}
\begin{eqnarray}
 \label{eq:mom_pi}
 \rho v_{\varpi} \frac{\partial v_{\varpi}}{\partial \varpi} &+& \rho
 v_{z} \frac{\partial v_{\varpi}}{\partial z} - \frac{\rho
 v_{\varphi}^{2}}{\varpi} \nonumber \\
 &=& - \rho \frac{\partial \psi}{\partial
 \varpi} - \frac{\partial p_{\rm tot}}{\partial \varpi}
- \frac{\langle B_{\varphi}^2 \rangle}{4 \pi \varpi } ~, 
\end{eqnarray}
\begin{eqnarray}
 \label{eq:mom_phi}
 \rho v_{\varpi} \frac{\partial v_{\varphi}}{\partial \varpi} + \rho
 v_{z} \frac{\partial v_{\varphi}}{\partial z} + \frac{\rho
 v_{\varpi} v_{\varphi}} {\varpi}
 = \nonumber \\
\frac{1}{{\varpi}^{2}}
 \frac{\partial}{\partial \varpi} \left[ {\varpi}^{2}
\frac{ \langle B_{\varpi} B_{\varphi} \rangle}{4\pi}
 \right] + \frac{\partial}{\partial z}
 \left( \frac{\langle B_{\varphi} B_{z} \rangle}{4 \pi}\right)~,
\end{eqnarray}
\begin{equation}
 \label{eq:mom_z}
 0 = - \frac{\partial \psi}{\partial z}
 - \frac{1}{\rho} \frac{\partial p_{\rm tot}}{\partial z}
 ~,
\end{equation}
\begin{eqnarray}
 \label{eq:ene}
 \frac{\partial}{\partial \varpi} \left[ \left( \rho \epsilon +
					   p_{\rm gas} + p_{\rm rad}\right)
 v_{\varpi}\right]
 + \frac{v_{\varpi}}{\varpi} \left( \rho \epsilon +
 p_{\rm gas} + p_{\rm rad} \right) \nonumber \\
+
 \frac{\partial}{\partial z} \left[
 \left( \rho \epsilon + p_{\rm gas} + p_{\rm rad} \right) v_{z}\right]
 - v_{\varpi} \frac{\partial}{\partial \varpi} \left( p_{\rm gas} + p_{\rm
 rad} \right) \nonumber \\
- v_{z} \frac{\partial}{\partial z} \left( p_{\rm gas} + p_{\rm rad} \right)
= q^{+} - q^{-} ~,
\end{eqnarray}
\begin{eqnarray}
 \label{eq:ind_phi}
 0 = -\frac{\partial}{\partial z} \left[ v_{z} \langle B_{\varphi} \rangle
 \right] -\frac{\partial}{\partial \varpi} \left[ v_{\varpi} \langle
 B_{\varphi} \rangle \right] \nonumber \\
+ \{ \nabla \times \langle \delta
  \mbox{\boldmath{$v$}} \times \delta \mbox{\boldmath{$B$}} \rangle
 \}_{\varphi} - \{ \eta \nabla \times \left( \nabla \times
  \mbox{\boldmath{$\bar{B}$}} \right) \}_{\varphi} ~,
\end{eqnarray}
where $p_{\rm tot} = p_{\rm gas} + p_{\rm rad} + p_{\rm mag}$ is the total
pressure and $ p_{\rm mag} = \langle B_{\varphi}^{2} \rangle /8 \pi$ is the
azimuthally averaged magnetic pressure. The third and fourth terms on the
right-hand side of equation (\ref{eq:ind_phi}) represent the dynamo term
and the magnetic
diffusion term which we approximate later on the basis of
the results of the
numerical simulations. Note that we have considered only the azimuthal component
of the induction equation (\ref{eq:vec_ind}).

\subsubsection{Vertically Integrated, Azimuthally Averaged Equations}

We assume that the radial velocity $v_{\varpi}$, the specific angular momentum
$\ell = \varpi v_{\varphi}$, and $\beta \equiv (p_{\rm gas} + p_{\rm
 rad})/p_{\rm mag}$ are independent of $z$, and we assume a polytropic
 relation $p_{\rm tot} = K \rho^{1+1/N}$ in the vertical
direction, where $N$ is the polytropic index (we adopt $N=3$ in this
paper). Vertical integration of equation (\ref{eq:mom_z}) yields
\begin{eqnarray}
 \label{eq:poly_rho}
 \rho(\varpi, z) = \rho_0 (\varpi)\left(1-\frac{z^2}{H^2}\right)^N ~,
\end{eqnarray}
\begin{eqnarray}
 \label{eq:poly_p}
 p_{\rm tot}(\varpi, z) = p_{{\rm tot} 0}(\varpi) \left( 1- \frac{z^2}{H^2}
 \right)^{N+1} ~,
\end{eqnarray}
and we assume 
\begin{eqnarray}
 \label{eq:poly_t}
 T(\varpi, z) = T_0(\varpi) \left(1-\frac{z^2}{H^2}\right) ~, 
\end{eqnarray}
where $H$ is the half thickness of the disk given by
\begin{eqnarray}
 \label{eq:h}
 \Omega_{{\rm K} 0}^2 H^2 = 2(N+1) \frac{p_{{\rm tot} 0} }{\rho_0} ~,
\end{eqnarray}
where $\Omega_{{\rm K}0}=(GM/\varpi)^{1/2}/(\varpi -r_{\rm s})$ is the Keplerian
angular velocity. Here the subscript $0$ refers to quantities in the
equatorial plane. Under these assumptions, the surface density $\Sigma$ and the vertically integrated
total pressure $W_{\rm tot}$ are given by
\begin{eqnarray}
 \label{eq:si}
 \Sigma &\equiv& \int_{-H}^{H}\rho dz \nonumber \\
&=&
 \int_{-H}^{H}\rho_0\left(1-\frac{z^2}{H^2}\right)^N dz \nonumber \\
&=& 2\rho_0I_NH ~,
\end{eqnarray}
\begin{eqnarray}
 \label{eq:wtot}
 W_{\rm tot} &\equiv& \int_{-H}^{H} p_{\rm tot}dz \nonumber \\
&=& \int_{-H}^{H}p_{{\rm tot}
  0} \left(1-\frac{z^2}{H^2} \right)^{N+1} dz \nonumber \\
&=& 2p_{{\rm tot} 0}I_{N+1}H ~,
\end{eqnarray}
where $I_N=(2^N N!)/(2N+1)!$. Using $\Sigma$ and $W_{\rm tot}$, equation
(\ref{eq:h}) can be rewritten as
\begin{eqnarray}
 \label{eq:h2}
 \Omega_{{\rm K} 0}^2 H^2 = (2N+3) \frac{W_{\rm tot} }{\Sigma} ~.
\end{eqnarray}
We note that the half thickness of the disk depends on not only the gas
and radiation pressure but also on the magnetic pressure.

We now integrate the other basic equations in the vertical direction and
assume Keplerian rotation ($\Omega = \Omega_{\rm K0}$) instead of
equation (\ref{eq:mom_pi}). We obtain
\begin{eqnarray}
 \label{eq:con_int}
 \dot{M} = -2\pi\varpi\Sigma v_\varpi
\end{eqnarray}
\begin{eqnarray}
 \label{eq:mom_pi_int}
 \Omega = \Omega_{\rm K0}
\end{eqnarray}
\begin{eqnarray}
 \label{eq:mom_phi_int}
 \dot M(\ell_{\rm K0}-\ell_{\rm in})= -2\pi \varpi^2 \int_{-H}^{H} \frac{\langle
 B_{\varpi} B_{\varphi} \rangle }{4 \pi} dz
\end{eqnarray}
\begin{eqnarray}
 \label{eq:ene_int}
 \frac{\dot M}{2\pi \varpi^2}\frac{W_{\rm gas}+W_{\rm rad}}{\Sigma} \xi = Q^+
 -Q_{\rm rad}^{-} ~,
\end{eqnarray}
\begin{eqnarray}
 \label{eq:ind_int}
 \dot \Phi &\equiv& \int_{-H}^{H}v_{\varpi} \langle B_{\varphi} \rangle
 dz \nonumber \\
&=& \int_{\varpi}^{\varpi_{\rm out}} \int_{-H}^{H}
 [ \{ \nabla \times \langle \delta
  \mbox{\boldmath{$v$}} \times \delta \mbox{\boldmath{$B$}} \rangle
 \}_{\varphi} \nonumber \\
&-& \{ \eta \nabla \times \left( \nabla \times
  \mbox{\boldmath{$\bar{B}$}} \right) \}_{\varphi} ] d\varpi dz +
\mbox{const.} 
\end{eqnarray}
where $\dot M$ is the mass accretion rate, $\ell_{{\rm K} 0} = \varpi^2
\Omega_{{\rm K} 0}$ is the Keplerian angular momentum and $\ell_{\rm in}$ is the
specific angular momentum swallowed by the black hole. $W_{\rm gas}$ and
$W_{\rm rad}$ are the vertically integrated gas and radiation pressures,
respectively. The left-hand side of the energy equation has been simplified by
introducing the entropy gradient parameter $\xi$ \citep[see][]{kato08}. 
On the right-hand side, $Q^{+}$ is the vertically integrated heating
rate, and $Q_{\rm rad}^{-}$ is the vertically integrated radiative cooling
rate. In equation (\ref{eq:ind_int}), $\dot \Phi$ is the radial
advection rate of the
azimuthal magnetic flux (hereafter we call it the magnetic flux advection
rate).

\subsection{$\alpha$-prescription of the Maxwell Stress Tensor}

Global MHD simulations of radiatively inefficient, magnetic pressure supported
(low-$\beta$) disks \citep{mach06} showed that the ratio of the azimuthally
averaged Maxwell stress to the sum of the azimuthally averaged gas
pressure and magnetic pressure is
nearly constant except in the plunging region ($\alpha_{\rm B} 
\equiv - \langle B_{\varpi} B_{\varphi} / 4 \pi \rangle / \langle
p_{\rm gas} + p_{\rm mag}
\rangle \sim 0.05 - 0.1$). Global MHD simulations of
radiatively inefficient, gas pressure supported 
(high-$\beta$) disks also implied that $\alpha_{\rm B}$ is nearly
constant because $\alpha_{\rm m}
\equiv - \langle B_{\varpi} B_{\varphi} / 4 \pi \rangle / \langle p_{\rm mag}
\rangle \sim 0.2 - 0.5$ and $\beta \sim 10$ are nearly constant inside
the disk \citep[e.g.,][]{hawl01}.

Following the simulation results, 
we assume
that the azimuthally
averaged $\varpi \varphi$-component of the Maxwell stress inside a disk is
proportional to the total (gas, radiation, and magnetic) pressure
\begin{eqnarray}
 \label{eq:al}
\frac{\langle B_{\varpi} B_{\varphi} \rangle }{4 \pi} = - \alpha p_{\rm tot}~.
\end{eqnarray}
Integrating in the vertical direction, we obtain
\begin{eqnarray} 
 \label{eq:al_int}
\int_{-H}^{H} \frac{\langle B_{\varpi} B_{\varphi} \rangle }{4 \pi} dz= -
\alpha W_{\rm tot}~.
\end{eqnarray}
This is one of the key assumptions in this paper. When the magnetic
pressure is high,
the stress level can be high even though the gas pressure and the radiation
pressure are low.

\subsection{Heating and Cooling Rates}
\subsubsection{The Magnetic Heating Rate}
In the conventional theory, the viscous heating was
expressed as $q^{+}_{\rm vis} = t_{\varpi \varphi} \varpi \left( d \Omega / d
\varpi \right)$, where $t_{\varpi \varphi}$ is the $\varpi
\varphi$-component of the total stress and $\Omega$ is the angular
velocity,
respectively. The results of three-dimensional MHD
simulations indicate that dissipation due to the thermalization of
magnetic energy dominates the total dissipative heating rate throughout
a disk and is expressed as $q^{+} \sim \langle B_{\varpi} B_{\varphi}
/ 4 \pi \rangle \varpi \left( d \Omega / d \varpi \right)$
\citep[e.g.][]{hiro06,mach06,krol07}. Hereafter, we refer to it as the magnetic
heating rate.

Following these simulation results, we employ magnetic heating as
the heating mechanism inside a disk, and set the vertically integrated
heating rate as follows:

\begin{eqnarray}
 \label{eq:qmag}
 Q^{+} = \int^{H}_{-H} \left[ \frac{\langle B_{\varpi}B_{\varphi}
 \rangle}{4\pi} \varpi
 \frac{d \Omega}{d \varpi} \right] dz = - \alpha W_{\rm tot} \varpi
 \frac{d \Omega}{d \varpi} ~ ,
\end{eqnarray}
where we have used equation (\ref{eq:al}). 
We note that if the magnetic pressure is high, the heating rate can also be
large even when the gas pressure and the radiation pressure are low.

\subsubsection{The Radiative Cooling Rate}
We assume that radiative cooling is provided by the thermal
bremsstrahlung emission in the optically thin limit. Thus the vertically
integrated optically thin cooling rate is given by
\begin{eqnarray}
 \label{eq:thin}
 Q^{-}_{\rm thin} &=& \int^{H}_{-H} 6.2 \times 10^{20} \rho^{2}
				 T^{1/2} dz \nonumber \\
 &=& 6.2 \times
 10^{20} \frac{I_{2N+1/2}}{2{I_N}^2} \frac{\Sigma^2}{H} {T_0}^{1/2} ~.
\end{eqnarray} 

In addition, we assume radiative cooling due to blackbody
radiation in the optically 
thick limit. The vertically integrated optically thick cooling rate
is expressed as

\begin{eqnarray}
 \label{eq:qthick}
 Q^{-}_{\rm thick} = \frac{16 \sigma I_{N} T_0^4}{3\tau / 2} ~,
\end{eqnarray}
where $\sigma$ is the Stefan-Boltzmann constant, $\tau = \tau_{\rm abs}
+ \tau_{\rm es}$ is the total optical depth, $\tau_{\rm abs}$ is the
absorption optical depth, $\tau_{\rm es} = 0.5 \kappa_{\rm es} \Sigma$
is the electron scattering optical depth, and $\kappa_{\rm es} = 0.34
 ~ {\rm cm}^2 ~ {\rm g}^{-1}$
is the electron scattering opacity. 

Although we should solve the transfer equation in the intermediate case,
we use the following approximate form \cite[e.g.,][]{hube90, nara95,
abra96}, 
\begin{eqnarray}
 \label{eq:qrad}
 Q^{-} = \frac{16 \sigma I_{N} {T_0}^4}{3\tau/2 + \sqrt{3} +
 {\tau_{\rm abs}}^{-1}} ~, 
\end{eqnarray}
where
\begin{eqnarray}
 \label{eq:tauabs}
 \tau_{\rm abs} = \frac{Q^{-}_{\rm thin}}{Q^{-}_{\rm thick}} =
 \frac{6.2 \times 10^{20}}{16 \sigma} \frac{I_{2N+1/2}}{2{I_N}^3}
 \frac{\Sigma^3}{H} {T_0}^{-7/2} ~.
\end{eqnarray}
Using these approximations, the vertically integrated radiation pressure
and the total pressure can be expressed as
\begin{eqnarray}
 \label{eq:wrad}
 W_{\rm rad} = \frac{Q^{-}}{4c} \frac{I_4}{I_N} H \left[ \frac{2}{3}
						  \left(
						  \frac{3\tau}{2} + \sqrt{3
						 }\right)\right] ~,
\end{eqnarray}

\begin{eqnarray}
 \label{eq:eos}
 W_{\rm tot} &=& W_{\rm gas}+W_{\rm rad}+W_{\rm mag} \nonumber \\
 &=& (1+\beta^{-1}) \left[\frac{I_{N+1}}{I_N} \frac{\mathcal{R}}{\mu} T_0
 \Sigma \right.] \nonumber\\
 &+& \left.[\frac{Q^{-}}{4 c} \frac{I_4}{I_N} H \left[
 \frac{2}{3}\left( \frac{3 \tau}{2} + \sqrt{3} \right)\right] \right] ~, 
\end{eqnarray}
respectively.

\subsection{Prescription of the Magnetic Flux Advection Rate}
We complete the set of basic equations by specifying the radial distribution of the
mean azimuthal magnetic fields. If we ignore the dynamo term and the
magnetic diffusion term and perform the integration in the left-hand side of the
induction equation (\ref{eq:ind_int}), we obtain
\begin{eqnarray}
 \label{eq:ind_int_ignore}
 \dot \Phi = -v_{\varpi} B_0(\varpi) 2 \frac{I_{(N+1)/2}}{\sqrt{I_N}}H =
 \mbox{const.}
\end{eqnarray}
where
\begin{eqnarray}
 \label{eq:b0}
 B_0(\varpi) = 2 \pi^{1/2} \left(\frac{\mathcal{R}}{\mu}T_0\right)^{1/2}\Sigma^{1/2}H^{-1/2}\beta^{-1/2}
\end{eqnarray}
is the mean azimuthal magnetic field in the equatorial
plane. However, $\dot \Phi$ is not always conserved in the radial direction
due to the presence of the dynamo term and the magnetic diffusion
term. According
to \cite{mach06}, this equals $\dot \Phi \propto \varpi^{-1}$ in the quasi steady
state. Following this result, we parametrize the
dependence of $\dot \Phi$ on $\varpi$ by introducing a parameter, $\zeta$, as follows. 
\begin{eqnarray}
 \label{eq:phidot}
 \dot \Phi(\varpi; \zeta, \dot M) \equiv {\dot \Phi}_{\rm out}(\dot M) \left(
  \frac{\varpi}{\varpi_{\rm out}}\right)^{-\zeta} ~,
\end{eqnarray}
where ${\dot \Phi}_{\rm out}$ is the magnetic flux advection rate at the
outer boundary
$\varpi = \varpi_{\rm out}$. When $\zeta=0$, the magnetic flux
is conserved in the radial direction, and when $\zeta>0$ (or $\zeta < 0$), the
magnetic flux
increases (or decreases) with a decreasing radius (see panel (b)
in figure \ref{line}). The azimuthal magnetic flux can increase
inside a disk when the azimuthal flux of opposite polarity buoyantly
escapes from the disk \citep[e.g.,][]{nish06}

Equation (\ref{eq:phidot}) is the second key assumption in this
paper. Specifying the magnetic flux advection rate enables the magnetic
pressure to increase when the disk temperature and the disk thickness
decrease. By contrast, if we
specified the plasma $\beta$ at each radius instead of the magnetic flux
advection rate, the decrease in
the temperature results in a decrease in 
magnetic pressure. This is inconsistent with the results of
three-dimensional MHD simulations \citep[e.g.,][]{mach06}.

\section{Results} \label{result}
\subsection{Local Thermal Equilibrium Curves} \label{res_local}

We solved the above basic equations at $\varpi = 5 r_{\rm s}$ to obtain
local thermal equilibrium curves.
We also assume $\ell_{\rm in} = \ell_{\rm K}(3r_{\rm s})$ in the $\varphi$
component of the
momentum equation (\ref{eq:mom_phi_int}). Results of the global
three-dimensional MHD simulations of optically thin black hole accretion
flows \cite[e.g.,][]{mach04} indicate that $\xi = 1$ \cite[see
also][]{oda07}. This positivity of $\xi$ means that the heat advection
becomes a cooling at a fixed radius. In this paper, we adopt $\xi = 1$ in
both the optically thin regime and the optically thick regime.

To determine the flux advection rate at a fixed radius, we need to specify
${\dot \Phi}_{\rm out}$. Hence, we imposed the outer boundary condition
that $T_{\rm out} = T_{\rm virial} = \left( \mu c^2 / 3 \mathcal{R}
\right) \left(\varpi_{\rm out}
/ r_{\rm s} \right)^{-1}$ and $\beta_{\rm out} = 10$ at
$\varpi_{\rm out}=1000r_{\rm s}$. Under these
conditions, we obtained ${\dot \Phi}_{\rm out}$ by solving
equations (\ref{eq:mom_phi_int}), (\ref{eq:al_int}), (\ref{eq:eos}), and
(\ref{eq:phidot}).
For given $\dot M$, $\alpha$, $\zeta$, and $\varpi$, we 
obtained local thermal equilibrium solutions.

\begin{figure}
 \epsscale{0.5}
 \plotone{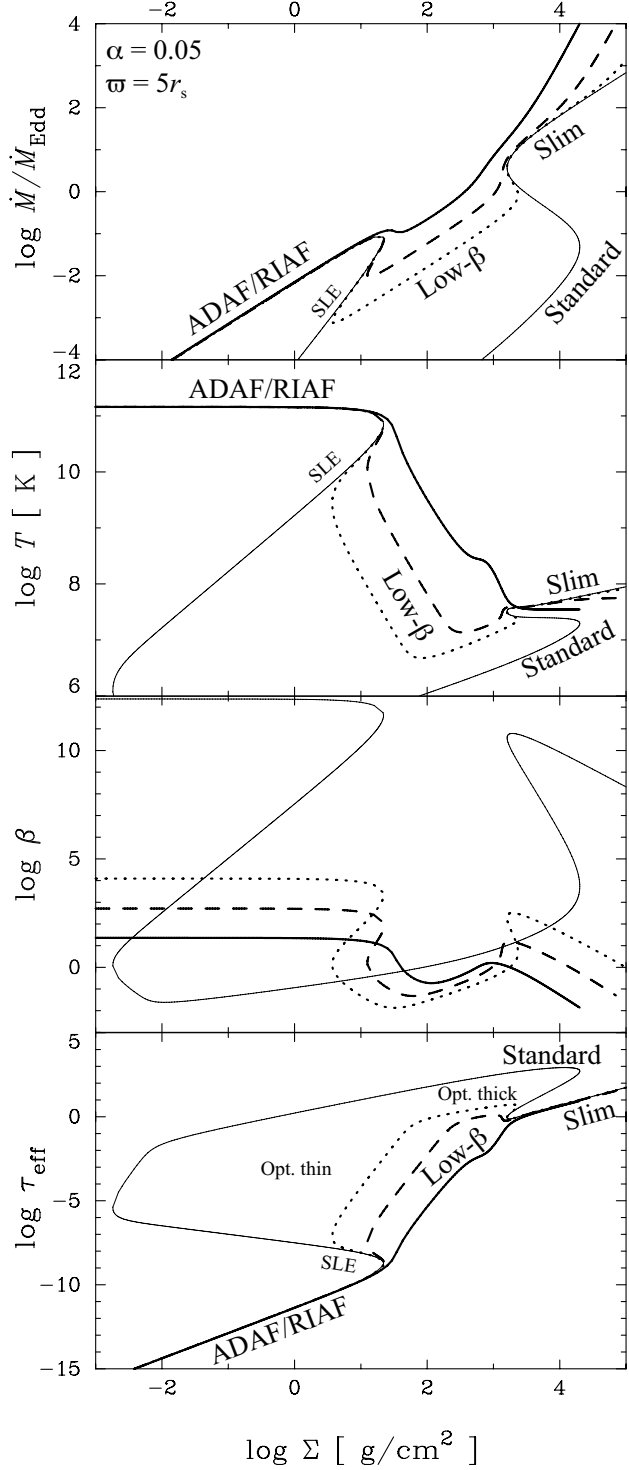}
 \caption{Local thermal equilibrium curves of accretion disks at
 $\varpi = 5 r_{\rm s}$ on the $\Sigma$ vs. ${\dot M}/{\dot M}_{\rm
 Edd}$, $T$, $\beta$ and $\tau_{\rm eff}$ plane for $\alpha = 0.05$, $\zeta
 = 0.6$(thick solid), $0.3$(dashed), $0$(dotted), and $-1.8$(thin
 solid, extremely weak magnetic field case). ${\dot M}_{\rm Edd} = L_{\rm Edd}/\eta_{\rm e} c^2$ is the
 Eddington accretion rate, where the energy conversion efficiency is
 taken to be $\eta_{\rm e} = 0.1$. The low-$\beta$ branches connect the
 optically
 thin and thick regimes. \label{x005al05sifoo}}
\end{figure}

Figure \ref{x005al05sifoo} shows the thermal equilibrium curves
plotted in the $\Sigma$ vs. $\dot M / {\dot M}_{\rm Edd}$, $T_{0}$, $\beta$,
$\tau_{\rm eff}$ plane for $\alpha = 0.05$, $\zeta = 0.6$ (thick solid),
$0.3$ (dashed), 
$0$ (dotted), and $-1.8$ (thin solid), respectively. Here ${\dot M}_{\rm
Edd} = L_{\rm Edd} / \eta_{\rm e} c^2 = 4 \pi GM / \left( \eta_{\rm e}
\kappa_{\rm es} c \right)$ is the
Eddington mass accretion rate, $\eta_{\rm e} = 0.1$ is the energy
conversion efficiency, $\tau_{\rm eff} =
0.5 \sqrt{\kappa_{\rm es} \kappa_{\rm ff}} ~\Sigma$ is the effective
optical depth, and $\kappa_{\rm ff} = 6.4 \times 10^{22} \rho_0
{T_0}^{-3.5} ~ {\rm cm}^2 ~ {\rm g}^{-1}$ is the opacity of free-free
absorption, respectively.

When $\zeta$ has a large negative value ($\zeta =
-1.8$), which means that the magnetic field strength at $\varpi = 5
r_{\rm s}$ is negligible, the ADAF/RIAF and
SLE branches appear in the optically thin regime, and the standard and slim disk
branches appear in the optically thick regime as expected. (Strictly
speaking, although the low-$\beta$ branches discussed below also appear
in much lower temperature regions, we do not display them because we
assumed fully ionized plasmas.)

For large values of $\zeta$, we obtained new branches in both the optically thin
and the optically thick
regime. We call them ``low-$\beta$'' branches. On the low-$\beta$
branches, the dissipated magnetic energy is mainly radiated rather than
advected. Therefore, the disk temperature and the gas pressure are lower than
in the ADAF/RIAF 
(but higher than in the standard disk). In contrast, the magnetic
pressure is high because the mean
azimuthal magnetic field increases as $H$ decreases.
The low-$\beta$ branches consist of an optically thin part and
an optically thick part.
We find that the optically thin low-$\beta$ branches
extend to $\dot M \gtrsim 0.1 {\dot M}_{\rm Edd}$ corresponding to $L
\gtrsim 0.1 L_{\rm Edd}$ when $\zeta \gtrsim 0.3$. We note that the
critical mass accretion rate for the existence of ADAF/RIAF
solutions ${\dot M}_{\rm c,A}$ is $\sim 0.1 {\dot M}_{\rm Edd}$
and that of the gas pressure dominated standard disk solutions ${\dot
M}_{\rm c,S}$ is $\sim 0.05 {\dot M}_{\rm Edd}$. 
The slim disk solution appears when $\dot M \gtrsim 5 {\dot M}_{\rm
Edd}$.

It is worth noting that $\dot M \propto \Sigma$ and $T_{0} \propto
\Sigma^{-2}$ in the optically thin low-$\beta$ branches, and that $\dot M
\propto \Sigma$ and $T_{0} \propto \Sigma^{1/2}$ in the optically thick
low-$\beta$ branches. These relations are derived as follows. 
First let us derive the
dependence of ${\dot
\Phi}_{\rm out} (\dot M)$ on $\dot M$. In the range of $\dot M \lesssim
{\dot M}_{\rm Edd}$ in which the low-$\beta$ branches appear, $W_{\rm
gas} \gg W_{\rm rad}$ at $\varpi_{\rm out}$. From equations
(\ref{eq:h2}), (\ref{eq:con_int}), (\ref{eq:mom_phi_int}), and
(\ref{eq:eos}), we find that $W_{\rm tot, out} \propto \dot M$,
$\Sigma_{\rm out} \propto \dot M$, $H_{\rm out} \sim {\rm constant}$,
and $v_{\varpi {\rm , out}} \sim {\rm constant}$. Therefore equation
(\ref{eq:b0}) yields $B_0(\varpi_{\rm out}) \propto {\dot M}^{1/2}$ and
equation (\ref{eq:ind_int_ignore}) yields ${\dot \Phi}_{\rm out} \propto
{\dot M}^{1/2}$. Next, we derive the dependence of $\dot M$ on
$\Sigma$. When the magnetic pressure dominates the total pressure,
$W_{\rm tot} \sim {B_0}^2 H \propto {\dot \Phi}^2 / ({v_\varpi}^2 H)
\propto {\dot M} / ({v_\varpi}^2 H)$. Since equations
(\ref{eq:con_int}), (\ref{eq:mom_phi_int}), and (\ref{eq:h2}) yield
$v_{\varpi} \propto \dot M / \Sigma$, $W_{\rm tot} \propto \dot M$, $H
\propto (W_{\rm tot} / \Sigma)^{1/2} \propto (\dot M / \Sigma)^{1/2}$,
we find that $\dot M \propto \Sigma$ and $H$ is constant. 
Finally we derive the dependence of $T_{0}$ on $\Sigma$ by noting that magnetic
heating ($Q^{+} \propto W_{\rm tot} \propto \dot M \propto \Sigma$)
balances radiative cooling on the low-$\beta$ branches. We find from
$Q^{-} \propto \Sigma^2 {T_{0}}^{1/2}$ that $T_{0} \propto \Sigma^{-2}$ in the
optically thin regime, whereas we find from $Q^{-} \propto
{T_{0}}^4/\Sigma$ that $T_{0} \propto \Sigma^{1/2}$  in the optically thick
regime.

\begin{figure}
 \epsscale{1.0}
 \plotone{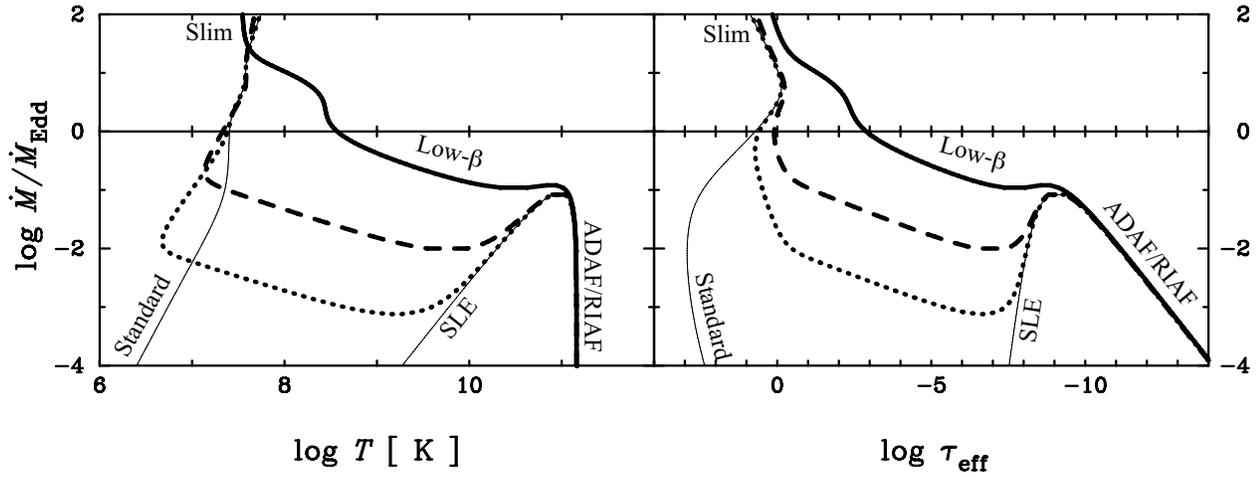}
 \caption{Same as figure \ref{x005al05sifoo} but plotted in the $T$ vs. ${\dot
 M}/{\dot M}_{\rm Edd}$ plane and the $\tau_{\rm eff}$ vs. ${\dot
 M}/{\dot M}_{\rm Edd}$ plane. The temperature $T$
 anti-correlates with $\dot M$ on the optically thin low-$\beta$
 branches. \label{x005al05mdfoo}}
\end{figure}

Figure \ref{x005al05mdfoo} depicts thermal equilibrium curves for the
same parameters as in figure \ref{x005al05sifoo} but plotted in the
$T_{0}$ vs. $\dot M / {\dot M}_{\rm Edd}$ plane and the $\tau_{\rm eff}$
vs. $\dot M / {\dot M}_{\rm Edd}$ plane. On the
ADAF/RIAF branch, the disk temperature is independent of
the mass accretion rate and is nearly constant at $T_{0}
\sim 10^{11} {\rm K}$. In contrast, on the
optically thin low-$\beta$ branches, the disk
temperature anti-correlates with the mass accretion rate, ${\dot M}
\propto T_{0}^{-1/2}$, and $T_{0} \sim 
10^{7-11} {\rm K}$. However, it is not possible to compare these results
directly with the observational data \citep[e.g., the cutoff energy
or the electron temperature in][]{miya08} because the disk temperature
does not represent the electron temperature. We will discuss this issue
in $\S$ \ref{dis_thin}.

\begin{figure}
 \epsscale{0.5}
 \plotone{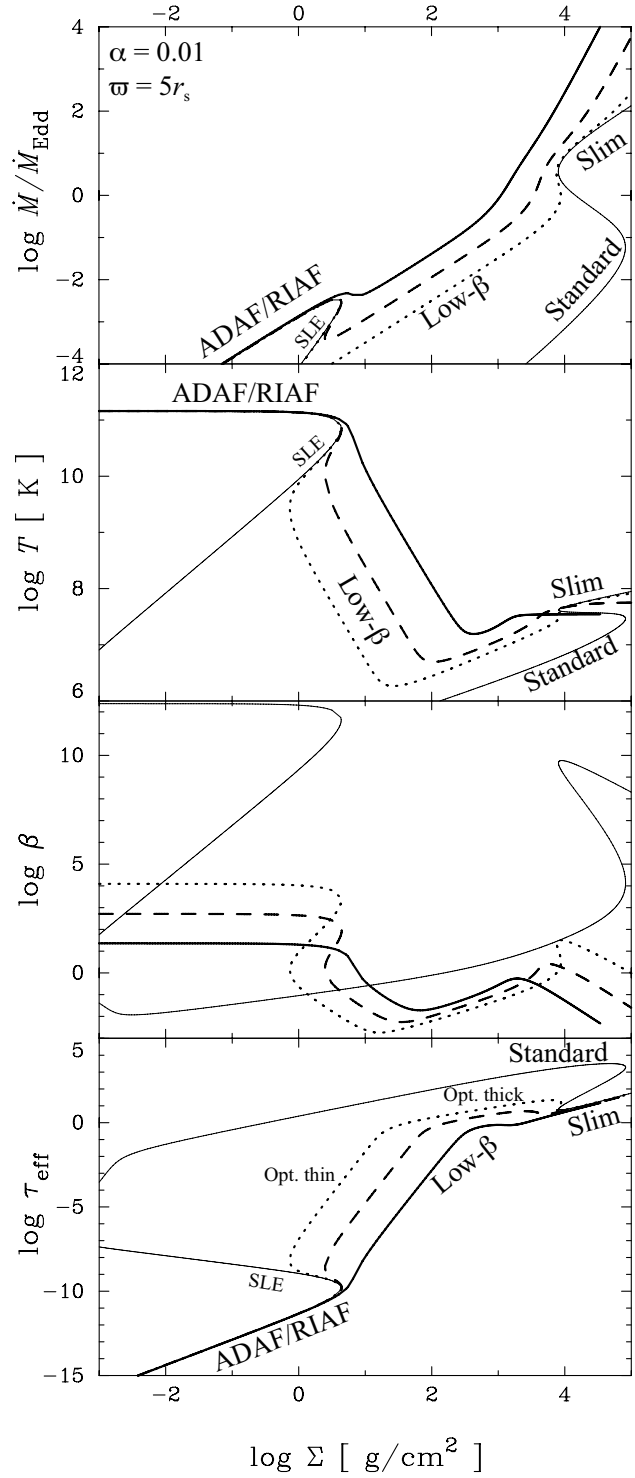}
 \caption{Same as figure \ref{x005al05sifoo}, but for $\alpha =
 0.01$. \label{x005al01sifoo}}
\end{figure}

Figure \ref{x005al01sifoo} shows the same thermal equilibrium curves as
in figure \ref{x005al05sifoo} but for $\alpha = 0.01$.
Compared to the results
for $\alpha = 0.05$, the ADAF/RIAF solution disappears 
at the lower mass accretion rate (${\dot M}_{\rm c,A} \sim 0.003
{\dot M}_{\rm Edd}$).

\subsection{The Radial Structure of the Disks at Thermal Equilibrium}
In the previous subsection, we obtained local thermal equilibrium curves
by solving the basic equations for various $\dot
M$ and fixed parameters, $\alpha$, $\zeta$, and $\varpi$. In this
subsection, we study the dependence of the solutions on $\varpi$ by
fixing the parameters, $\alpha$, $\zeta$, and $\dot M$.
Since we focus on the hard-to-soft transition of the
BHCs, we assume that a disk is in the
ADAF/RIAF state corresponding to the low/hard state when the mass accretion
rate is low. Thus we choose the ADAF/RIAF solution when three
solutions (ADAF/RIAF, SLE, and low-$\beta$ or standard disk solution)
are found for the same mass accretion rate.

\begin{figure}
 \epsscale{0.8}
 \plotone{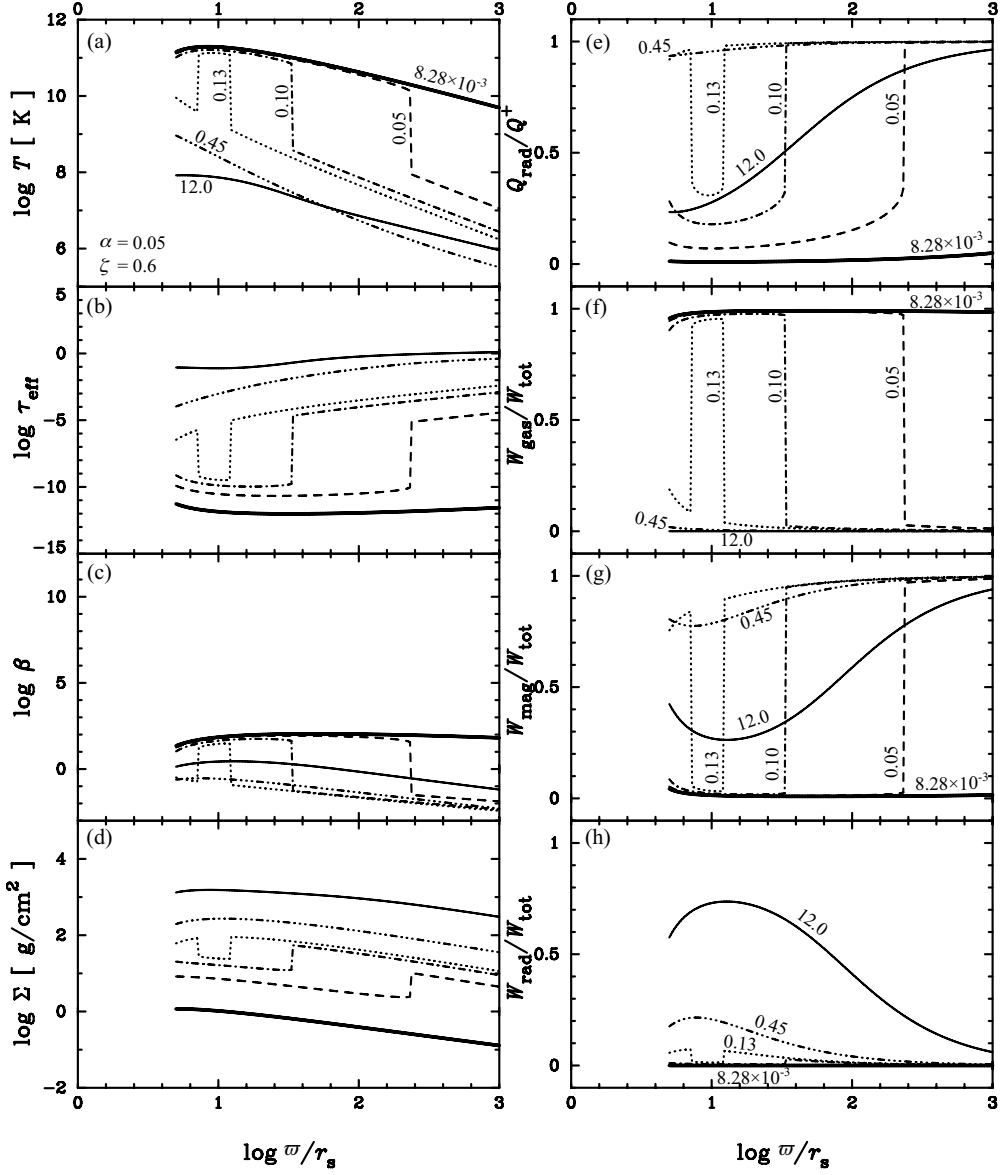}
 \caption{The radial dependence of (a), temperature $T$, (b)
 effective optical depth, $\tau_{\rm eff}$, (c) $\beta = (p_{\rm
 gas} + p_{\rm rad}) / p_{\rm mag}$, (d) surface density $\Sigma$, (e)
 the ratio of the heating rate and the radiative cooling rate,
 the ratio of the vertically integrated total pressure and (f) the gas
 pressure $W_{\rm gas}$, (g) the magnetic pressure $W_{\rm mag}$, and (h)
 the radiation pressure $W_{\rm rad}$ for $\alpha = 0.05$, $\zeta =
 0.6$, $\dot M/{\dot M}_{\rm Edd} = 8.28 \times 10^{-3}$(thick solid),
 $0.05$(dashed), $0.10$(dot-dashed), $0.13$(dotted),
 $0.45$(triple-dot-dashed), $12$(thin solid). We chose the ADAF/RIAF
 solution when we obtained three solutions (ADAF/RIAF, SLE and low-$\beta$
 or standard disk solutions) for the same $\dot M$. \label{al05zp06xfoo}}
\end{figure}

Figure \ref{al05zp06xfoo} shows the radial structure of a disk for $\alpha =
0.05$, $\zeta = 0.6$, $\dot M/ {\dot M}_{\rm Edd} = 8.28 \times
10^{-3}$(thick solid), $0.05$(dashed), $0.10$(dot-dashed),
$0.13$(dotted), $0.45$(triple-dot-dashed), and $12.0$(thin solid),
respectively.

When the
mass accretion rate is low ($\dot M \lesssim 0.01 {\dot
M}_{\rm Edd}$), the dissipated energy is advected
inward because radiative cooling is inefficient due to the low
surface density over the whole disk. The whole disk is hot and the gas pressure
is dominant, that is, the disk is in the ADAF/RIAF state.

As the mass accretion rate
increases ($0.01 {\dot M}_{\rm Edd} \lesssim
\dot M \lesssim 0.2 {\dot M}_{\rm Edd}$), the disk
undergoes a transition to the optically thin low-$\beta$ disk from the outer
radii because the mass accretion rate exceeds ${\dot M}_{\rm c,A}$
at the transition radius.
Now let us derive the radial dependence of ${\dot M}_{\rm c,A}$. We find
from the local thermal equilibrium curves that the
ADAF/RIAF branch in which the heating rate balances the
heat advection rate ($Q^{+} \sim Q_{\rm adv}$) intersects with the SLE branch
in which the heating rate balances the
optically thin radiative cooling rate ($Q^{+} \sim Q^{-}_{\rm
thin}$) at ${\dot M}_{\rm c,A}$. The gas pressure is
dominant in both branches ($W_{\rm tot} \sim W_{\rm gas}$). Using
these relations, we find from the basic equations that ${\dot M}_{\rm c,A}
\propto \alpha^{2}
\varpi^{5} \Omega_{\rm K0}^{-1} \left| d \Omega_{\rm K0} / d \varpi
\right|^{3} \left( \ell_{\rm K0} - \ell_{\rm in}\right)$. Since ${\dot
M}_{\rm c,A}$ increases inward in the outer region,
the transitions take place from the outer radius. We note that ${\dot
M}_{\rm c,A}$ has its maximum value around $10 r_{\rm s}$
 for $\ell_{\rm in} = \ell_{\rm K0} (3 r_{\rm s})$. Therefore, the 
innermost region ($\varpi
\lesssim 10 r_{\rm s}$) undergoes a transition to the optically thin
low-$\beta$ disk at a lower mass accretion rate than in 
the slightly further out region ($10 r_{\rm s} \lesssim \varpi \lesssim 30 r_{\rm
s}$).
If we consider
transonic solutions for which $\ell_{\rm in} < \ell_{\rm K0}$, such
a feature may not appear.

When $0.2 {\dot
M}_{\rm Edd} \lesssim
\dot M \lesssim 5 {\dot M}_{\rm Edd}$, the whole disk becomes
an optically thin low-$\beta$ disk in which the dissipated energy is
radiated by the bremsstrahlung emission. 
We note that when $\zeta = 0.6$, an
optically thick low-$\beta$ 
disk does not appear because the large magnetic flux prevents the disk
from shrinking in the vertical direction. 
Let us derive the radial dependence of the surface density and the
temperature in optically thin low-$\beta$ disks. 
In an optically thin low-$\beta$ disk, the magnetic pressure is
dominant ($W_{\rm tot} \sim W_{\rm mag}$)
and the magnetic heating rate balances the optically thin
radiative cooling rate ($Q^{+} \sim Q_{\rm thin}^{-}$). Using these
relations, we find from the basic equations that $\Sigma \propto \dot M
\alpha^{-1}
\varpi^{4\zeta/5 - 2} \left( \ell_{\rm K0} - \ell_{\rm in}\right)^{3/5}
\Omega_{\rm K0}^{-2/5}$ and $T_{0} \propto {\dot M}^{-2} \alpha^{4}
\varpi^{6-4\zeta} \left( d \Omega_{\rm K0} / d \varpi \right)^{2}
$. 
In the outer region where $\Omega_{\rm K0} \propto \varpi^{-3/2}$ and
$\left( \ell_{\rm K0} - \ell_{\rm in}\right) \sim \ell_{\rm K0} \propto
\varpi^{1/2}$, we find that $ \Sigma \propto \dot M \alpha^{-1}
\varpi^{4\zeta/5-11/10}$ and $T_{0} \propto {\dot M}^{-2} \alpha^{4}
\varpi^{1-4\zeta}$.

When the mass
accretion rate is high ($\dot M \gtrsim 5 {\dot
M}_{\rm Edd}$), the disk undergoes a transition to a slim disk in the 
inner region because the radial heat advection becomes efficient
again. We note that even
in a slim disk the magnetic pressure is comparable to or larger than
the radiation pressure ($W_{\rm mag} \gtrsim W_{\rm rad} \gg W_{\rm
gas}$) for $\zeta = 0.6$.

\begin{figure}
 \epsscale{0.8}
 \plotone{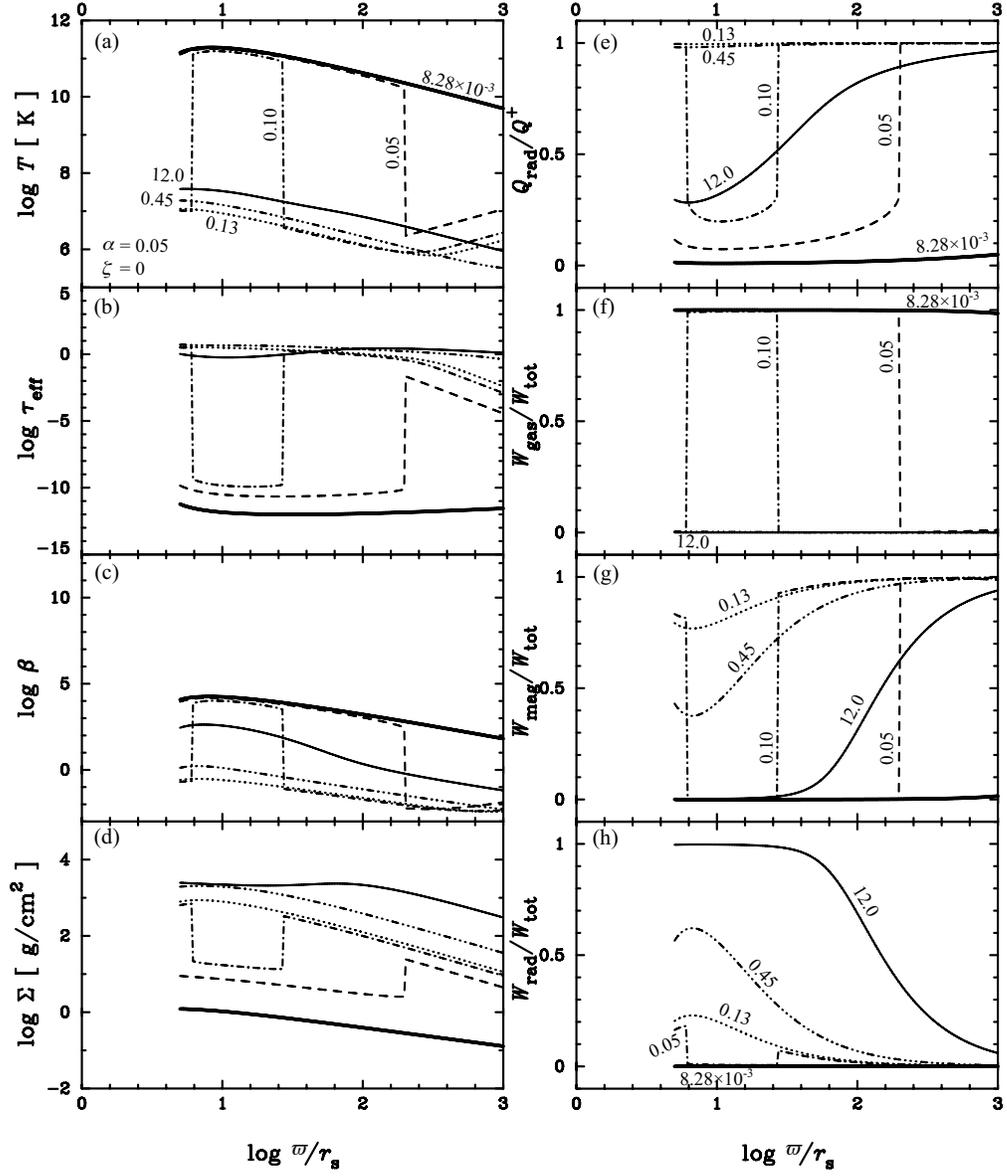}
 \caption{Same as figure \ref{al05zp06xfoo}, but for $\zeta =
 0$. \label{al05z000xfoo}}
\end{figure}

Figure \ref{al05z000xfoo} shows the results for $\zeta = 0$. As the mass
accretion rate increases ($0.01 {\dot M}_{\rm Edd} \lesssim
\dot M \lesssim 0.2 {\dot M}_{\rm Edd}$), the disk undergoes a transition to
a low-$\beta$ disk from the outer radii, and the whole disk becomes a
low-$\beta$ disk when $0.2 {\dot M}_{\rm Edd} \lesssim
\dot M \lesssim 5 {\dot M}_{\rm Edd}$. The low-$\beta$ disks consist of
two parts, an 
inner optically thick part and an outer optically thin part. In an
optically thick low-$\beta$ disk, since $Q^{-} \sim Q_{\rm thick}^{-}
\propto 16 \sigma I_{N} {T_{0}}^4 / \left(3 \tau / 2 \right) \propto
{T_{\rm eff}}^{4}$, we obtain the radial dependence of the effective
temperature, $T_{\rm eff} \propto \varpi^{-3/4}$. This dependence is
the same as that in the standard disk. In
addition, when the mass accretion rate is high ($\dot M \gtrsim 5 {\dot
M}_{\rm Edd}$), slim disks consist of an inner
radiation pressure dominated region and an outer magnetic pressure
dominated region.

\begin{figure}
 \epsscale{0.8}
 \plotone{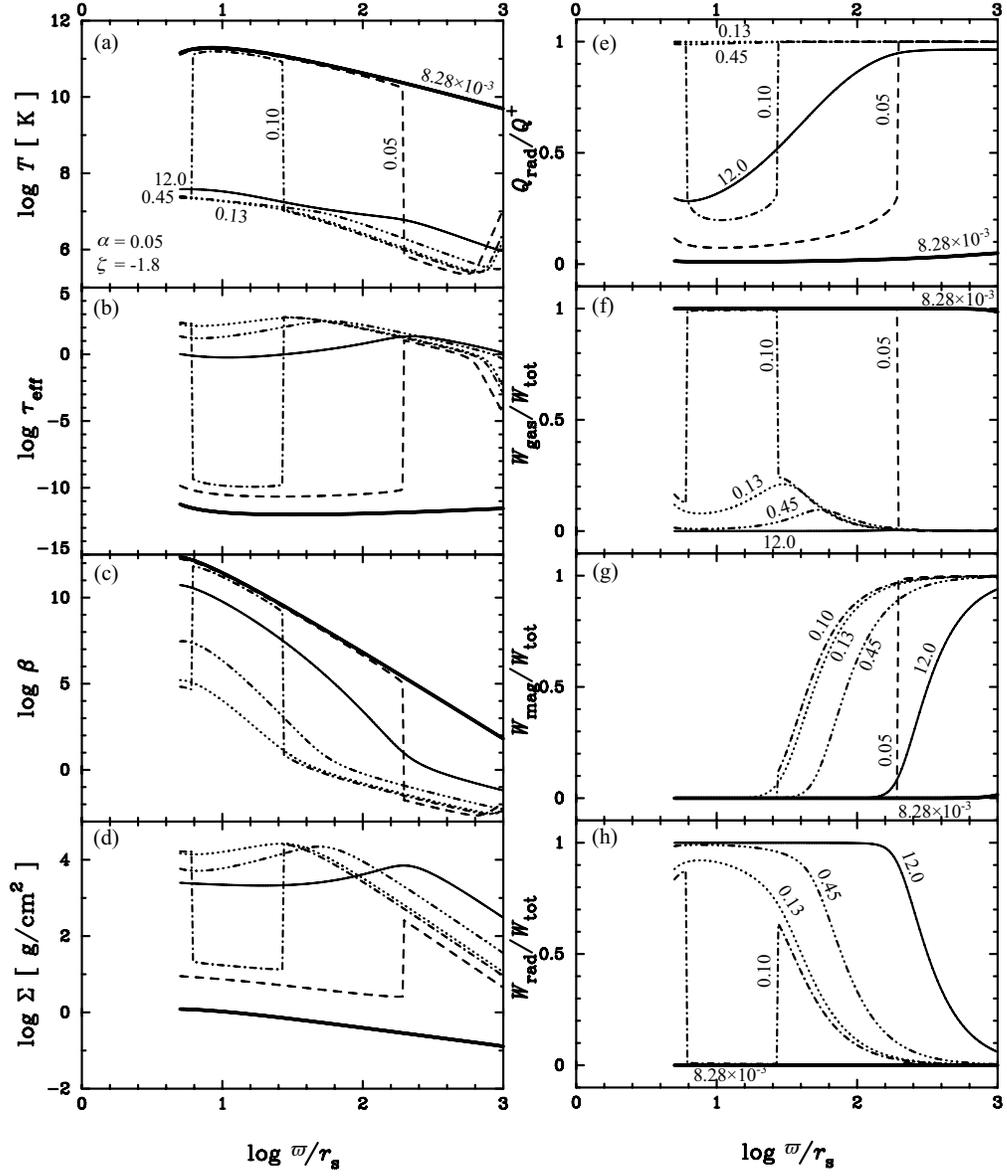}
 \caption{Same as figure \ref{al05zp06xfoo}, but for $\zeta = -1.8$.
 \label{al05zm18xfoo}}
\end{figure}

An example of an extremely weak magnetic field model ($\zeta = -1.8$) is
shown in figure
\ref{al05zm18xfoo}. In this model, the inner region of a disk undergoes
a transition to a radiation pressure dominated standard disk while the
outer region undergoes a transition to an optically thick low-$\beta$
disk. As a result, the inner region of a disk always stays in the high
$\beta$ state.

\begin{figure}
 \epsscale{0.8}
 \plotone{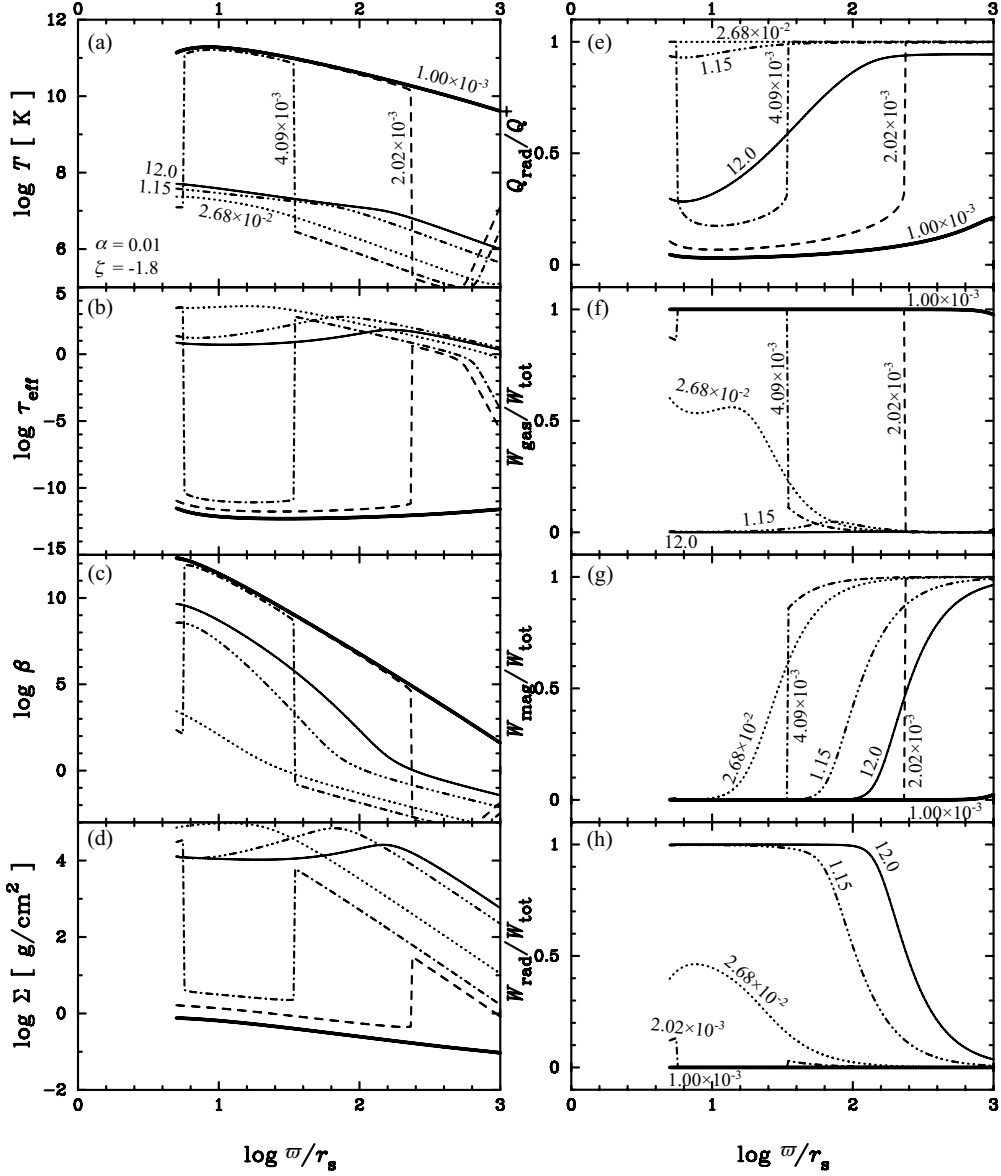}
 \caption{Same as figure \ref{al05zp06xfoo}, but for $\alpha = 0.01$,
 $\zeta = -1.8$, $\dot M/{\dot M}_{\rm Edd} = 1.00 \times 10^{-3}$(thick solid),
 $2.02 \times 10^{-3}$(dashed), $4.09 \times 10^{-3}$(dot-dashed), $2.68
 \times 10^{-2}$(dotted),
 $0.45$(triple-dot-dashed), $12$(thin solid). \label{al01zm18xfoo}}
\end{figure}

When $\alpha = 0.05$, the gas pressure dominated
standard disk solutions do not appear because ${\dot M}_{\rm c,A}$,
being proportional to $\alpha^2$, is
higher than ${\dot M}_{\rm c,S}$
for $\alpha = 0.05$. The transition to the gas pressure dominated
standard disk takes place when $\alpha$ is sufficiently small so that
${\dot M}_{\rm c,A} < {\dot M}_{\rm c,S}$. In figure \ref{al01zm18xfoo},
we show the results 
for $\alpha = 0.01$, $\zeta=-1.8$, $\dot M/ {\dot M}_{\rm Edd} = 1 \times
10^{-3}$(thick solid), $2.02 \times
10^{-3}$(dashed),
$4.09 \times 10^{-3}$(dot-dashed), $2.68 \times 10^{-2}$(dotted),
$1.15$(triple-dot-dashed), and $12.0$(thin solid), respectively. As the mass
accretion rate increases ($1 \times 10^{-3} {\dot M}_{\rm Edd} \lesssim
\dot M \lesssim 0.03 {\dot M}_{\rm Edd}$), the disk
undergoes a transition to a gas pressure dominated standard disk in the
inner region while an optically thick low-$\beta$ disk exists in the outer
region. 

\section{Discussion} \label{discussion}
\subsection{New Thermal Equilibrium Solution connecting an Optically Thin Regime
 and an Optically Thick Regime}
We obtained thermal equilibrium curves for optically thin and
optically thick accretion
disks incorporating the mean azimuthal magnetic fields. We
prescribed that the azimuthally averaged Maxwell stress is proportional
to the total (gas, radiation, and magnetic) pressure. Consequently, the
heating rate can be large
if the magnetic pressure is high even when the gas pressure and
the radiation pressure are low. To complete the set of basic equations,
we specified the magnetic flux advection rate rather than the plasma
$\beta$ at each radius. Hence, a decrease in
temperature via the cooling instability results in an increase in 
magnetic pressure due to the conservation of the azimuthal magnetic flux
$\langle B_{\varphi} \rangle H$.

Under these assumptions, we obtained 
low-$\beta$ solutions in both the optically thin regime and the optically
thick regime in addition to the ADAF/RIAF, SLE, standard, and slim disk
solutions. In the low-$\beta$ disks, the magnetic heating balances
the radiative cooling. 
The slopes of the thermal equilibrium curves plotted in
the $\Sigma$ vs. $T$ plane in figure \ref{x005al05sifoo} and figure 
\ref{x005al01sifoo} indicate that the low-$\beta$ disk is
thermally stable. 
The borders between the optically thin and thick parts of the low-$\beta$
branches depend on $\zeta$ which specifies the magnetic
flux at each radius. As the magnetic flux advection rate increases, the magnetic
pressure becomes higher, therefore the magnetic heating rate
increases. Hence the equilibrium temperature becomes higher, and the
surface density decreases. Thus, the effective optical depth
decreases for the same mass
accretion rate. This means that the optically thin part of the
low-$\beta$ branch extends to a higher mass accretion rate as $\zeta$
increases.

We also studied the radial dependence of the solutions. The ADAF/RIAF
undergoes a transition to the low-$\beta$ disk from the outer radii as the mass
accretion rate increases. This is because ${\dot M}_{\rm c,A}$ at
outer radii
 is lower than its value
at inner radii. Figure \ref{disks} shows schematic pictures of
configurations of accretion disks for various $\dot M$ and $\zeta$.

Let us discuss the stability of low-$\beta$ disks. \cite{shib90} carried
out MHD simulations of the buoyant escape
of the magnetic flux due to the Parker instability. 
They showed that once a disk is dominated by
the magnetic pressure, it can stay in the low-$\beta$ state because the strong
magnetic tension suppresses the growth of the Parker
instability. \cite{mine95} suggested that such low-$\beta$ disks emit
hard X-rays. When
$\beta \sim 1$ in the surface of the low-$\beta$ disk, the buoyant
escape of the magnetic flux from the surface layer will reduce the total
azimuthal magnetic flux of the disk. This loss of magnetic flux can
balance the amplification of magnetic fields due to the MRI in
this layer. The results of three-dimensional global MHD simulations by
\cite{mach06} indicate that low-$\beta$ disks stay in a
quasi-equilibrium state at least for a thermal time scale. The longer time
scale evolution of the low-$\beta$ disk is a subject to be studied in
near future.

\subsection{Bright Hard State undergoing Bright/Slow Transition}
\label{dis_thin}
One of the main aims of this paper is to explain the bright hard state observed
in rising phases of the outbursts of BHCs \citep[e.g.,
][]{miya08}.
Figure \ref{x005al05mdfoo} shows that the
optically thin part of the low-$\beta$ branch
extends to $\gtrsim 0.1 {\dot M}_{\rm
Edd}$ for $\zeta \gtrsim 0.3$. In the ADAF/RIAF branches, the disk
temperature is 
independent of the mass accretion rate and remains nearly at the virial
temperature. This is consistent with the fact that the cutoff energy observed
in the low/hard state of BHCs is independent of their luminosity. In
contrast, the disk temperature roughly
anti-correlates with the mass accretion rate in the optically thin low-$\beta$
branches. This indicates that the electron temperature in the hard X-ray emitting
optically thin low-$\beta$ disks anti-correlates with the
luminosity. This feature is consistent with the anti-correlation
observed in the bright hard state of BHCs. 

In addition, the optically thin
low-$\beta$ disks explain the bright/slow transition reported by
\cite{gier06}. For small $\zeta$, the ADAF/RIAF directly undergoes a transition to
the optically thick disks emitting the soft X-rays at $\sim 0.1
{\dot M}_{\rm Edd}$ for $\alpha = 0.05$ (panel (h) $\rightarrow$ (g)
$\rightarrow$ (f) in figure \ref{disks}).
This corresponds to the dark/fast transition
\citep[see][]{gier06}. In contrast, for large $\zeta$, the ADAF/RIAF first
undergoes a transition to the
optically thin low-$\beta$ disk, which can be identified as the
bright hard state. Subsequently, this disk
undergoes a transition to the optically thick disk at a mass
accretion rate higher than the rate for small $\zeta$. This transition corresponds
to the bright/slow transition. (panel (h) $\rightarrow$
(j) $\rightarrow$ (i) $\rightarrow$ (e) in figure \ref{disks})

It is anticipated that if the magnetic flux escapes from the disk due to 
magnetic buoyancy
or dissipates due to magnetic reconnection, the equilibrium curve may
shift to the one for smaller $\zeta$. In such a case, the optically thin
low-$\beta$ disk immediately undergoes a transition to the
standard disk or the optically thick low-$\beta$ disk. This transition
may accompany the ejection of jets observed
during the hard-to-soft transition of BHCs because magnetic energy
stored in the low-$\beta$ disk is released. 

Although we have adopted the single temperature model
($T_{\rm i} = T_{\rm e}$), it
has been pointed out that the electron temperature becomes lower
than the ion temperature ($T_{\rm i} \sim
10^{11-12} {\rm K}$) in such a low density, high temperature
region \citep[e.g.,][\citeyear{naka97}]{naka96}. Moreover, cooling via the
synchrotron emission and the inverse-Compton effect can be efficient. By
contrast, the electron temperature and the ion temperature coincide
($T_{\rm i} \sim T_{\rm e} \sim 10^{6-7} {\rm K}$) in low
temperature and high density disks such as optically thick low-$\beta$
disks, standard disks, and slim disks.
In subsequent papers, we will take into account the decoupling of ions
and electrons in the low density, high temperature region and incorporate
the synchrotron emission and the inverse-Compton
effect, in addition to the bremsstrahlung emission as the cooling mechanism.

\subsection{Magnetically Supported Thermal Disks} \label{dis_thick}
We obtained solutions for the optically thick low-$\beta$ disks emitting 
blackbody radiation (panel (f) in figure \ref{disks}). The radial distribution
of the effective temperature in the optically thick low-$\beta$
disk is roughly the same as that for the standard disk ($T_{\rm
eff} \propto \varpi^{-3/4}$) because it does not depend on whether the gas
pressure or the magnetic pressure is dominant as long as the heating
rate balances the blackbody cooling. This
indicates that the optically thick low-$\beta$ disk will be observed as a
thermal disk.

The seed photons incident on the Compton corona (the low density, high
temperature region, that is, the ADAF/RIAF or the optically thin
low-$\beta$ region) have been thought
to be provided by a cool disk or by synchrotron emissions. 
The optically thick low-$\beta$ disk can also be the source of
seed photons. When $\zeta
\lesssim 0$, the ADAF/RIAF undergoes a transition to the optically thick
low-$\beta$ disk from the radially outer region as the mass accretion
rate increases.
Thus, the inner ADAF/RIAF region is surrounded by the outer optically
thick low-$\beta$ disk (panel (g) in figure \ref{disks}). An increase
in seed photons can result in
a decrease in the temperature of the Compton corona. 
Furthermore, the hard X-rays Comptonized in the
corona can be incident on the optically thick low-$\beta$ disk and can be
partly reflected. As a result, the existence of such region will enhance the
Compton continua with a reflection component in the X-ray spectrum.

To sum up, the optically thick low-$\beta$ disk not only
provides seed photons to the Compton corona but also produces a soft
spectral excess and a reflection component observed in the low/hard
state of Cyg X-1 \citep{maki08} and GRO J1655-40 \citep{taka08} at the
Suzaku X-ray observatory.

\subsection{Limit Cycle Oscillation Between an Optically Thick Low-$\beta$ Disk and a
 Slim Disk}
\begin{figure}
 \epsscale{0.8}
 \plotone{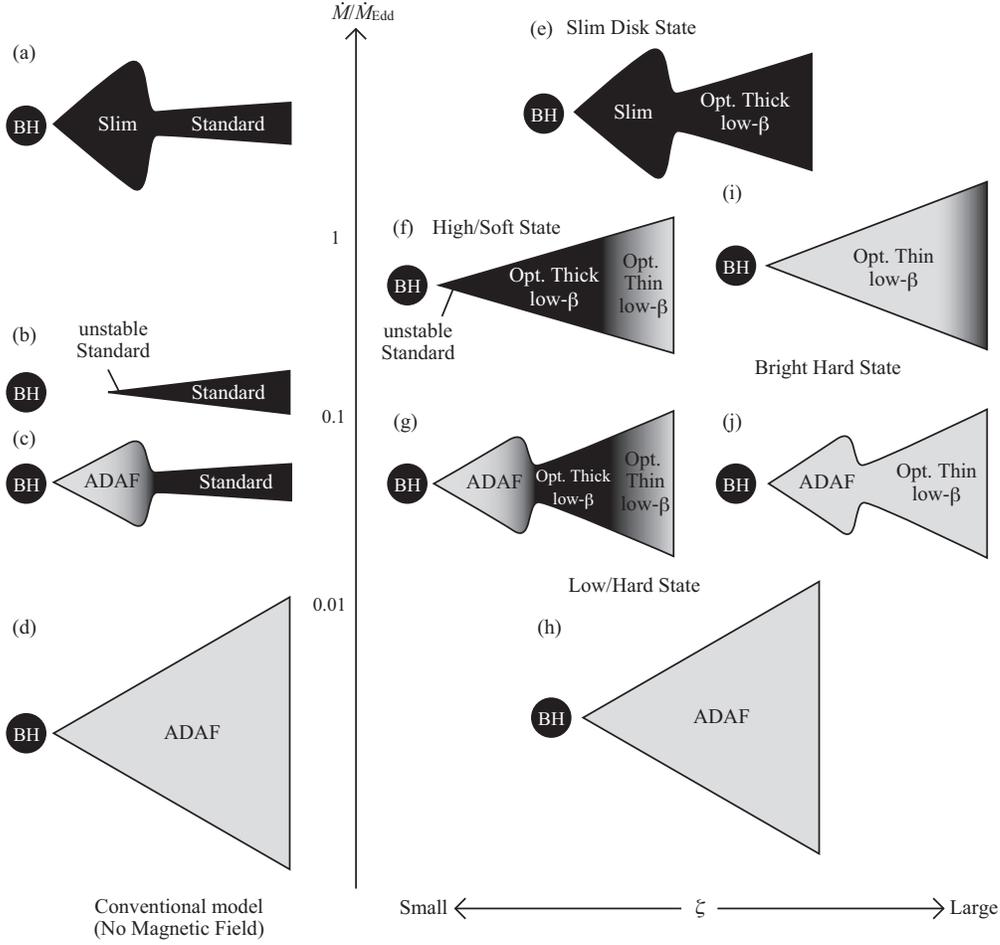}
 \caption{Schematic pictures of the configuration of accretion disks
 for various mass accretion rates and magnetic flux advection
 parameter $\zeta$. The optically
 thin region is indicated by gray, and the optically thick region is
 indicated by black. The panels (c), (d), (g), and (h) represent
 low mass accretion states corresponding to the
 low/hard state. The panels (i) and (j) represent mildly high mass
 accretion states for large $\zeta$ corresponding to the
 bright hard state. The panels (b) and (f) represent the mildly high mass
 accretion state for small $\zeta$ corresponding to the high/soft
 state. The slim disk state is illustrated in the top panels (a) and
 (e). The left column shows the 
 configuration predicted by the conventional (no magnetic) model. \label{disks}}
\end{figure}

So far, we have focused on the hard-to-soft transition of BHCs. 
Now we discuss X-ray flares as observed in GRS
1915+105 \citep[e.g.,][]{bell97,taam97,paul98,bell00}. This object
frequently exhibits state transitions and 
quasi-regular X-ray flares. It has been thought that the latter effect may be driven by
limit cycle oscillations between the standard disk and the slim disk
(panel (a) $\leftrightarrow$
(b) in figure \ref{disks}). \cite{ohsu06,ohsu07} reproduced such limit
cycle evolution by global two-dimensional radiation hydrodynamic (RHD)
simulations of
optically thick accretion disks using the flux-limited diffusion
\citep[FLD;][]{leve81}
approximation for the radiation transfer and the phenomenological
$\alpha$ viscosity for the angular momentum transfer and the viscous
heating. However, such RHD simulations do not follow the formation of the
low-$\beta$ disk presented in this paper.

We first consider a transition from
a slim disk to an optically thick low-$\beta$ disk. We suppose that
the disk is initially in the slim disk state in which magnetic
heating balances the heat advection (the photon trapping) and that these
are greater than the radiative cooling ($Q^{+} \sim Q_{\rm adv} > Q^{-}$)
. Figure \ref{x005al05sifoo} shows that as the mass accretion rate
decreases, the slim disk branches disappear. At this point, the magnetic
heating (and the heat advection) become less than the radiative
cooling ($Q^{+} \sim Q_{\rm adv} < Q^{-}$). Therefore cooling 
instability occurs. As a result, the disk shrinks in
the vertical direction along the cooling timescale and undergoes a transition
to an optically thick low-$\beta$ disk (panel (e) $\rightarrow$ (f) in
figure \ref{disks}). This situation is similar to
the transition from the ADAF/RIAF to the optically thin low-$\beta$
disk found by \cite{mach06}.
Subsequently, as the mass
accretion rate increases, the disk undergoes a transition back to a slim disk
because the optically thick low-$\beta$ branches
disappear (panel (f) $\rightarrow$ (e) in figure \ref{disks}).

We emphasize that an optically thick low-$\beta$ disk undergoes a transition
to a slim disk at higher mass accretion rates ($\sim
{\dot M}_{\rm Edd}$ for $\alpha = 0.5$ and $\zeta = 0$) than
those at which a standard disk undergoes a transition to a slim disk
($\sim 0.1 {\dot M}_{\rm Edd}$). This results in a small variation
in luminosity for quasi-regular X-ray flares
compared to that expected from the limit cycle between a standard disk
and a slim disk.

\section{Summary} \label{summary}

We have obtained low-$\beta$ solutions for optically thin and
optically thick disks incorporating the mean azimuthal magnetic
fields. The key assumptions are (1) we prescribed the 
Maxwell stress to be proportional to the total (gas, radiation, and
magnetic) pressure, and (2) we specified the magnetic flux advection
rate instead of the plasma $\beta$ at each radius. 
As a result, the strong magnetic fields mainly contribute to the heating
and increase the half thickness of the disk. In the low-$\beta$ disks,
the heating
enhanced by strong magnetic pressure balances the radiative cooling
rather than the heat advection. Optically thin low-$\beta$ disks
explain the bright hard state undergoing the bright/slow transition
observed in BHCs. The
optically thick low-$\beta$ disk can be observed as a thermal
disk, which can supply seed 
photons into the Compton corona (the ADAF/RIAF or the optically thin
low-$\beta$ region) and reflect incident photons from
the Compton corona. We pointed out the possibility of new limit cycle
oscillations between an optically
thick low-$\beta$ disk and a slim disk, which will exhibit a smaller variation
in luminosity than that expected from a limit cycle between a
standard disk and a slim disk.

\acknowledgments
We are grateful to T. Hanawa, S. Mineshige for discussion. We also
acknowledge T. Miyakawa for valuable comments and
discussions from observational points of view, and B. Prager for
helpful comments on the manuscript. This work is supported by
the Grant-in-Aid for Scientific Research of the Ministry of Education,
Culture, Sports, Science and Technology (RM: 20340040) and
Research Fellowships of Japan Society for the Promotion of Science for
Young Scientists.






\appendix


\onecolumn

\clearpage 
\clearpage

\clearpage

\clearpage

\clearpage

\clearpage

\clearpage

\clearpage

\clearpage

\clearpage



\begin{thebibliography}{}
\bibitem[Abramowicz et al.(1996)]{abra96}
		Abramowicz, M. A., Chen, X. M., Granath, M., \& Lasota,
		J. P. 1996, \apj, 471, 762

\bibitem[Abramowicz et al.(1995)]{abra95} 
 Abramowicz, M. A., Chen, X., Kato, S., Lasota, J. P., \& Regev, O. 1995,
 \apj, 438, L37

\bibitem[Abramowicz et al.(1988)]{abra88}
 Abramowicz, M. A., Czerny, B., Lasota, J. P., \& Szuszkiewicz,
 E. 1988, \apj, 332, 646

\bibitem[Balbus \& Hawley(1991)]{balb91}
 Balbus, S. A. \& Hawley, J. F. 1991, \apj, 376, 214


\bibitem[Belloni et al.(2000)]{bell00}
 Belloni, T., Klein-Wolt, M., Me\'{n}dez, M., van der Klis, M. \& van
		Paradijs, J. 2000, \aap, 355, 271

\bibitem[Belloni et al.(1997)]{bell97}
 Belloni, T., Me\'{n}dez, M., King, A. R., van der Klis, M. \& van
		Paradijs, J. 1997, \apj, 479, L145

\bibitem[Done \& Gierli\'{n}ski(2003)]{done03}
 Done, C. \& Gierli\'{n}ski, M. 2003, \mnras, 342, 1041

\bibitem[Eardley, Lightman, \& Shapiro(1975)]{eard75}
 Eardley, D. M., Lightman, A. P., \& Shapiro, S. L. 1975, \apj, 199, L153

\bibitem[Esin et al.(1997)]{esin97}
 Esin, A., McClintock, J. E., Narayan, R. 1997, \apj, 489, 865

\bibitem[Gierli\'{n}ski \& Newton(2006)]{gier06}
 Gierli\'{n}ski, M. \& Newton, J. 2006, \mnras, 370, 837

\bibitem[Hawley(2000)]{hawl00}
 Hawley, J. F. 2000, \apj, 528, 462

\bibitem[Hawley \& Krolik(2001)]{hawl01}
 Hawley, J. F. \& Krolik, J. H. 2001, \apj, 548, 348

\bibitem[Hawley \& Krolik(2002)]{hawl02}
 Hawley, J. F. \& Krolik, J. H. 2002, \apj, 566, 164


\bibitem[Hirose et al.(2006)]{hiro06}
 Hirose, S., Krolik, J. H., \& Stone, J. M. 2006, \apj, 640, 901

\bibitem[Hubeny(1990)]{hube90}
		Hubeny, I. 1990, \apj, 351, 632

\bibitem[Ichimaru(1977)]{ichi77}
 Ichimaru, S. 1977, \apj, 214, 840

\bibitem[Kato et al.(2008)]{kato08}
 Kato, S., Fukue, J., \& Mineshige, S. 2008, Black-Hole Accretion Disks:
		Towards a New Paradigm (Kyoto: Kyoto University Press)

\bibitem[Kawaguchi(2003)]{kawa03}
 Kawaguchi, T. 2003, \apj, 593, 69

\bibitem[Komissarov(2006)]{komi06}
 Komissarov, S. S. 2006, \mnras, 368, 993

\bibitem[Krolik et al.(2007)]{krol07}
 Krolik, J. H., Hirose, S., \& Blaes, O. 2007, \apj, 664, 1045

\bibitem[Levermore \& Pomraning(1981)]{leve81}
 Levermore, C. D. \& Pomraning, G. C. 1981, \apj, 248, 321

\bibitem[Machida, Hayashi, \& Matsumoto(2000)]{mach00}
 Machida, M., Hayashi, M. R., \& Matsumoto, R. 2000, \apj, 532, L67

\bibitem[Machida \& Matsumoto(2003)]{mach03}
 Machida, M. \& Matsumoto, R. 2003, \apj, 585, 429

\bibitem[Machida et al.(2004)]{mach04}
		Machida, M., Nakamura, K. E., \& Matsumoto, R. 2004,
		\pasj, 56, 671

\bibitem[Machida et al.(2006)]{mach06}
 Machida, M., Nakamura, K. E., \& Matsumoto, R. 2006, \pasj, 58, 193

\bibitem[Makishima et al.(2008)]{maki08}
 Makishima, K., Takahashi, H., Yamada, S. et al. 2008, \pasj, 60, 585

\bibitem[Mineshige et al.(1995)]{mine95}
 Mineshige, S., Kusunose, M. \& Matsumoto, R. 1995, \apj, 445, L43

\bibitem[Miyakawa et al.(2008)]{miya08}
 Miyakawa, T., Yamaoka, K., Homan, J., Saito, K., Dotani, T.,
 Yoshida, A., \& Inoue, H 2008, \pasj, 60, 637

\bibitem[Nakamura et al.(1996)]{naka96}
 Nakamura, K. E., Matsumoto, R., Kusunose, M., \& Kato, S. 1996,
		\pasj, 48, 761

\bibitem[Nakamura et al.(1997)]{naka97}
 Nakamura, K. E., Kusunose, M., Matsumoto, R., \& Kato, S. 1997,
		\pasj, 49, 503

\bibitem[Narayan \& Yi(1994)]{nara94}
 Narayan, R. \& Yi, I. 1994, \apj, 428, L13
 
\bibitem[Narayan \& Yi(1995)]{nara95}
 Narayan, R. \& Yi, I. 1995, \apj, 452, 710

\bibitem[Nishikori, Machida \& Matsumoto(2006)]{nish06}
 Nishikori, H, Machida, M, \& Matsumoto, R. \apj, 2006, 641, 862

\bibitem[Oda et al.(2007)]{oda07}
 Oda, H., Machida, M., Nakamura, K. E., \& Matsumoto, R. 2007, \pasj,
 59, 457

\bibitem[Ohsuga (2006)]{ohsu06}
 Ohsuga, K. 2006, \apj, 640, 923

\bibitem[Ohsuga (2007)]{ohsu07}
 Ohsuga, K. 2007, \apj, 659, 205

\bibitem[Okada et al.(1989)]{okad89}
 Okada, R., Fukue, J. \& Matsumoto, R. 1986, \pasj, 41, 133

\bibitem[Paczy\'{n}sky \& Wiita(1980)]{pacz80}
 Paczy\'{n}sky, B. \& Wiita, P. J. 1980, \aap, 88, 23

\bibitem[Parker(1966)]{park66}
 Parker, E. N. 1966, \apj, 145, 811

\bibitem[Paul et al.(1998)]{paul98}
 Paul, B., Agrawal, P. C., Rao, A. R., Vahia, M. N., \& Yadav,
		J. S. 1998, \apj, 492, L63

\bibitem[Shakura \& Sunyaev(1973)]{shak73}
 Shakura, N. I. \& Sunyaev, R. A. 1973, \aap, 24, 337

\bibitem[Shapiro, Lightman, \& Eardley(1976)]{shap76}
 Shapiro, S. L., Lightman, A. P., \& Eardley, D. M. 1976, \apj, 204, 187

\bibitem[Shibata, Tajima, \& Matsumoto(1990)]{shib90}
 Shibata, K., Tajima, T., \& Matsumoto, R. 1990, \apj, 350, 295

\bibitem[Shibazaki \& H\=oshi(1975)]{shib75}
 Shibazaki, N. \& H\=oshi, R. 1975, Progress of Theoretical Physics, 54, 706

\bibitem[Taam, Chen, \& Swank(1997)]{taam97}
 Taam, R. E., Chen, X., \& Swank, J. H. 1997, \apj, 485, L83

\bibitem[Takahashi et al.(2008)]{taka08}
 Takahashi, H., Fukazawa, Y., Mizuno, T et al. 2008, \pasj, 60, S69

\bibitem[Thorne \& Price(1975)]{thor75}
 Thorne, K. S. \& Price, R. H. 1975, \apj, 195, L101

\bibitem[Vierdayanti et al.(2006)]{vier06}
 Vierdayanti, K., Mineshige, S., Ebisawa, K, \& Kawaguchi, T. 2006,
		\pasj, 58, 915

\bibitem[Vierdayanti et al.(2008)]{vier08}
 Vierdayanti, K., Watarai, K, \& Mineshige, S. 2008,
		\pasj, 60, 653


\end{thebibliography}
\end{document}